\documentclass[a4paper,11pt]{article}
\pdfoutput=1 

\usepackage{jheppub} 

\usepackage[T1]{fontenc} 
\usepackage{slashed}

\usepackage{multirow}

\usepackage{cleveref}
\usepackage[numbers,sort&compress]{natbib}
\usepackage{mathtools}
\usepackage{xspace}
\usepackage{xcolor}
\usepackage{multicol}
\usepackage{svg}
\usepackage{booktabs}
\newcommand{\bea} {\begin{eqnarray}}
\newcommand{\eea} {\end{eqnarray}}
\newcommand{\beq} {\begin{equation}}
\newcommand{\eeq} {\end{equation}}
\def\gev{~{\rm GeV}}

\newcommand{\code}{\texttt}

\usepackage{listings, dirtree, mdframed}
\definecolor{keyword}{HTML}{008000}
\definecolor{emph}{HTML}{0000FF}
\definecolor{string}{HTML}{A52A2A}
\definecolor{comment}{rgb}{0.0, 0.44, 1.0}
\definecolor{back}{HTML}{F8F8F8}
\definecolor{arrow}{HTML}{745334}

\allowdisplaybreaks 

\title{Enhancing Phase Transition Calculations with Fitting and Neural Network}
\author[b]{Ligong Bian,}
\author[c]{Hongxin Wang,}
\author[a]{Yang Xiao,} 
\author[d]{Ji-Chong Yang,}
\author[a,e]{Jin Min Yang,}
\author[a]{Yang Zhang}

\affiliation[a]{School of Physics, Henan Normal University, Xinxiang 453007, P. R. China}
\affiliation[b]{Department of Physics and Chongqing Key Laboratory for Strongly Coupled Physics, Chongqing University, Chongqing 401331, P. R. China}
\affiliation[c]{Schools of Physics, Shandong University, Jinan 250100, P. R. China}
\affiliation[d]{Department of Physics, Liaoning Normal University, Dalian 116029, P. R. China}
\affiliation[e]{Institute of Theoretical Physics,  Chinese Academy of Sciences, Beijing 100190, P. R. China}

\emailAdd{lgbycl@cqu.edu.cn}
\emailAdd{wanghx@mail.sdu.edu.cn}
\emailAdd{xiaoyangphy@gmail.com}
\emailAdd{yangjichong@lnnu.edu.cn}
\emailAdd{jmyang@itp.ac.cn}
\emailAdd{zhangyang2025@htu.edu.cn}

\abstract{
The computation of bounce action in a phase transition involves solving partial differential equations, inherently introducing non-negligible numerical uncertainty. Deriving characteristic temperatures and properties of this transition necessitates both differentiation and integration of the action, thereby exacerbating the uncertainty. 
In this work, we fit the action curve as a function of temperature to mitigate the uncertainties inherent in the calculation of the phase transition parameters.
We find that, after extracting a factor, the sixth-order polynomial yields an excellent fit for the action in the high temperature  approximated potential. 
In a realistic model, the singlet extension of the Standard Model, this method performs satisfactorily across most of the parameter space after trimming the fitting data.
This approach not only enhances the accuracy of phase transition calculations but also systematically reduces computation time and facilitates error estimation, particularly in models involving multiple scalar fields.
Furthermore, we discussed the possible of using multiple neural networks to predict the action curve from model parameters.
}

\begin{document}
\maketitle
\flushbottom

\section{Introduction}

The first-order electroweak phase transition (EWPT) is a key topic in particle cosmology~\cite{Athron:2023xlk,Zhou:2022mlz, Bian:2021dmp, Kawana:2022lba, Xiao:2022oaq, Ramsey-Musolf:2024zex, Carena:2022yvx, Fabian:2020hny}. In the Standard Model (SM), the 125 GeV Higgs mass lead the EWPT to be a crossover~\cite{kajantie1997non}. Thus, any experimental evidence indicating a first-order EWPT would suggest the presence of new physics. Experimentally, a first-order EWPT can be indirectly verified through phenomenological variables. In particle physics, a first-order EWPT provides a non-equilibrium environment that facilitates baryogenesis, fulfilling one of the Sakharov conditions~\cite{sakharov1998violation}. In cosmology, the collision of vacuum bubbles formed during the first-order EWPT is a potential source of stochastic gravitational waves at the electroweak scale~\cite{Kosowsky:1991ua, Kosowsky:1992rz, Kosowsky:1992vn}. Furthermore, interactions between these vacuum bubbles and the thermal fluid can generate sound waves and turbulence, which contribute significantly to the gravitational waves produced during high-temperature phase transitions~\cite{Hindmarsh:2013xza,Huber:2008hg, Hindmarsh:2015qta, Hindmarsh:2017gnf, cutting2020vorticity, Auclair:2022jod, Caprini:2009yp, Caprini:2019egz}. Since the first detection of gravitational waves by LIGO~\cite{LIGOScientific:2016aoc}, many space exploration missions have been proposed to search for traces of stochastic gravitational waves. The upcoming space-based detectors, such as LISA~\cite{amaro2017laser}, Taiji~\cite{10.1093/nsr/nwx116}, and Tianqin~\cite{TianQin:2015yph}, are anticipated to begin detecting stochastic gravitational waves in the next decade. Therefore, precise calculations of gravitational wave parameters are crucial for future experimental discoveries.

There are many studies that investigate the uncertainties in the effective potential~\cite{Athron:2022jyi, DiLuzio:2014bua, Braathen:2016cqe, Niemi:2020hto, Lofgren:2021ogg, Qin:2024dfp, Curtin:2016urg, Zhu:2025pht} and the uncertainties  arising from various approximations in the computational framework for phase transition gravitational waves~\cite{Guo:2023gwv, Sharma:2023mao, RoperPol:2023dzg, Cai:2023guc, Jinno:2024nwb, guo2021benefits,cutting2018gravitational,Caprini:2019egz, Caprini:2006jb,cutting2020vorticity}.
The uncertainties in the numerical calculation of the bounce action for phase transition have been noted in the manuals of numerical tools, but their influence has not been sufficiently studied.
The path deformation method for higher-dimensional potential used in \code{CosmoTransitions}~\cite{Wainwright:2011kj}, \code{BMSPT3}~\cite{Basler:2024aaf}, and \code{Phasetracer2}~\cite{Athron:2020sbe,Athron:2024xrh} inherently leads to non-negligible uncertainties. 
Specialized algorithms and tools, such as \code{AnyBubble}~\cite{Masoumi:2016wot}, 
\code{SimpleBounce}~\cite{Sato:2019wpo},
\code{FindBounce}~\cite{Guada:2020xnz} and \code{BubbleProfiler}~\cite{Athron:2019nbd}
have been developed to accurately calculate the action. However, achieving this improved accuracy incurs certain sacrifices, such as requiring an effective potential that can be formulated through an analytical expression. Small uncertainties in the action can propagate to significant uncertainties in its derivative. Unfortunately, the inverse phase transition duration $\beta$, a crucial parameter for gravitational wave calculations, is derived from the derivative of the action with respect to temperature.

Another major challenge in action calculation lies in the extensive computation time required. For instance, calculating a single action can consume several seconds for a two-dimension potential and even minutes for a three-dimension potential. Accurately determining the characteristic temperatures of the transition necessitates integrating the action with respect to temperature, a process that necessitates computing the action hundreds of times. 

These two issues hinder the investigation of EWPTs in models with multiple scalar fields. A practical remedy is nonlinear fitting: by fitting a smooth function to the action with uncertainties small compared to the temperature scale, we can reduce the uncertainties in its derivative and accelerate the integral evaluation using the fitted function, thereby saving significant computation time. The aim of this work is to explore the utilization of polynomial fitting in accelerating the computation of phase transition characteristics and mitigating the instability in Euclidean action calculations, thereby providing more accurate and reliable characteristic parameters for phase transitions.

The paper is structured as follows. Section 2 presents evidence of numerical instabilities in the calculation of the action and $\beta$. In Section 3, we demonstrate the feasibility of using polynomial fitting for the action function in both a simplified one-dimensional model and singlet extensions of the Standard Model~(SSM), while addressing challenges encountered during the fitting process. Section 4 shows that the action curve may be predicted from model parameters using neural networks. We conclude in Section 5 and the implementation of this method in \code{CosmoTransitions} and \code{PhaseTracer} is documented in the appendix.

\section{Challenges in action calculation}

A first-order EWPT is a tunneling process between two background vacuums, which is typically triggered by the formation of bubbles (spherical nucleation) within the false vacuum background. Due to the pressure difference between the inside and outside of the bubble, the bubble expands outward, gradually transforming the background vacuum of the entire universe into the true vacuum. In this process, the field configuration of the background scalar field that connects the false vacuum and the true vacuum $\phi(r=\sqrt{x^2+y^2+z^2})$ is determined by the following differential equation ~\cite{Callan:1977pt, Coleman:1977py, Linde:1981zj}
\begin{equation} \label{eq: bubble profile}
    \frac{{\rm d}^2 \phi}{{\rm d} r^2} + \frac{2}{r}\frac{{\rm d} \phi}{{\rm d}r} = \frac{\partial V_{\rm eff}(\phi;T)}{\partial \phi},
\end{equation}
subjecting to the boundary conditions $\lim \limits_{r \to \infty} \phi(r) = 0 $ and ${\rm d} \phi/{\rm d} r|_{r=0}=0$ (see Ref.\cite{Rubakov:2002fi} for details).

If only one scalar field acquires a vacuum expectation value, the above equation can be analogized to the motion of a small ball in a potential $-V_{\rm eff}(\phi; T)$, and it can be solved using the shooting method~\cite{Apreda:2001us}. However, for multiple scalar fields cases, the above equation becomes a system of differential equations. Solving this system becomes more challenging and requires more complex algorithms, such as the path deformation techniques~\cite{Wainwright:2011kj}. 

In the path deformation method, the desired path $\phi(r)$ in the field space has been reparameterized as $\Vec{\phi}\left[r(\chi)\right]$. Starting from the initially guessed path and the requirement of the derivative $\left|\frac{{\rm d} \Vec{\phi}}{{\rm d} \chi}\right| = 1$, the equation of motion can be split into two parts, one parallel and one perpendicular to the path:
\begin{align} \label{eq: bubble profile path}
    \frac{{\rm d}^2 \chi}{{\rm d} r^2} + \frac{2}{r}\frac{{\rm d} \chi}{{\rm d}r} &= \frac{\partial V_{\rm eff}\left[\Vec{\phi}(\chi);T\right]}{\partial \chi}, \notag \\
    \frac{{\rm d}^2 \Vec{\phi}}{{\rm d} \chi^2}\left(\frac{{\rm d}\chi}{{\rm d}r}\right)^2 &= \nabla_{\perp} V_{\rm eff}(\Vec{\phi};T),
\end{align}
where $\nabla_{\perp}V_{\rm eff}$ is the components of the gradient of $V_{\rm eff}$ that are perpendicular to the path. 
After obtaining the bubble configuration, the Euclidean action could be directly calculated via
\begin{equation}
     S_{E} =  4 \pi \int^{+\infty}_{0}r^2{\rm d}r~\left[\frac{1}{2}\left(\frac{\partial \phi}{\partial r}\right)^{2} + V_{\rm eff}(\phi;T)\right].
\end{equation}

\begin{figure}[thbp!]
    \centering
    \includegraphics[width=0.9\textwidth]{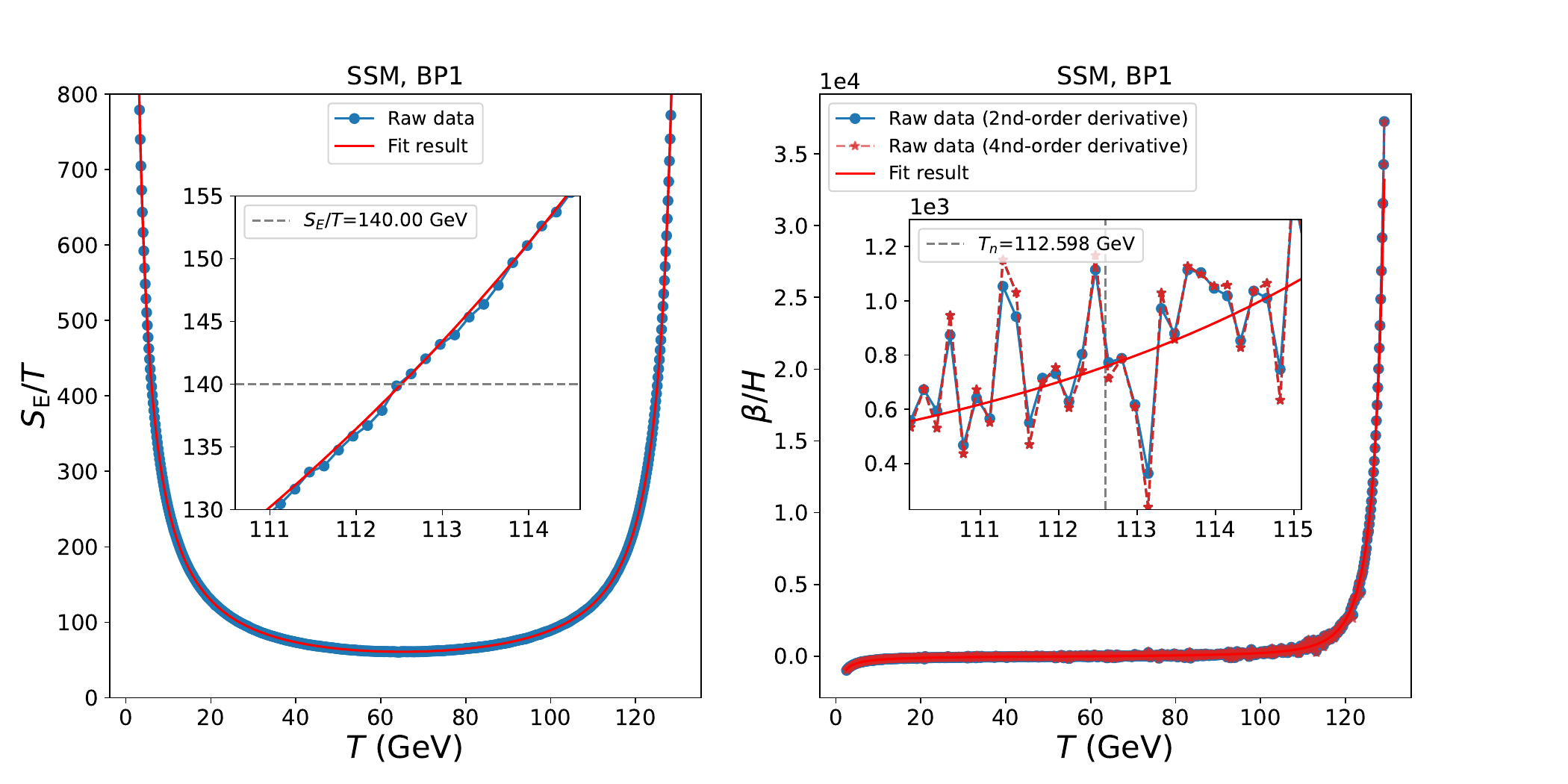}\\
    
    \includegraphics[width=0.9\textwidth]
    {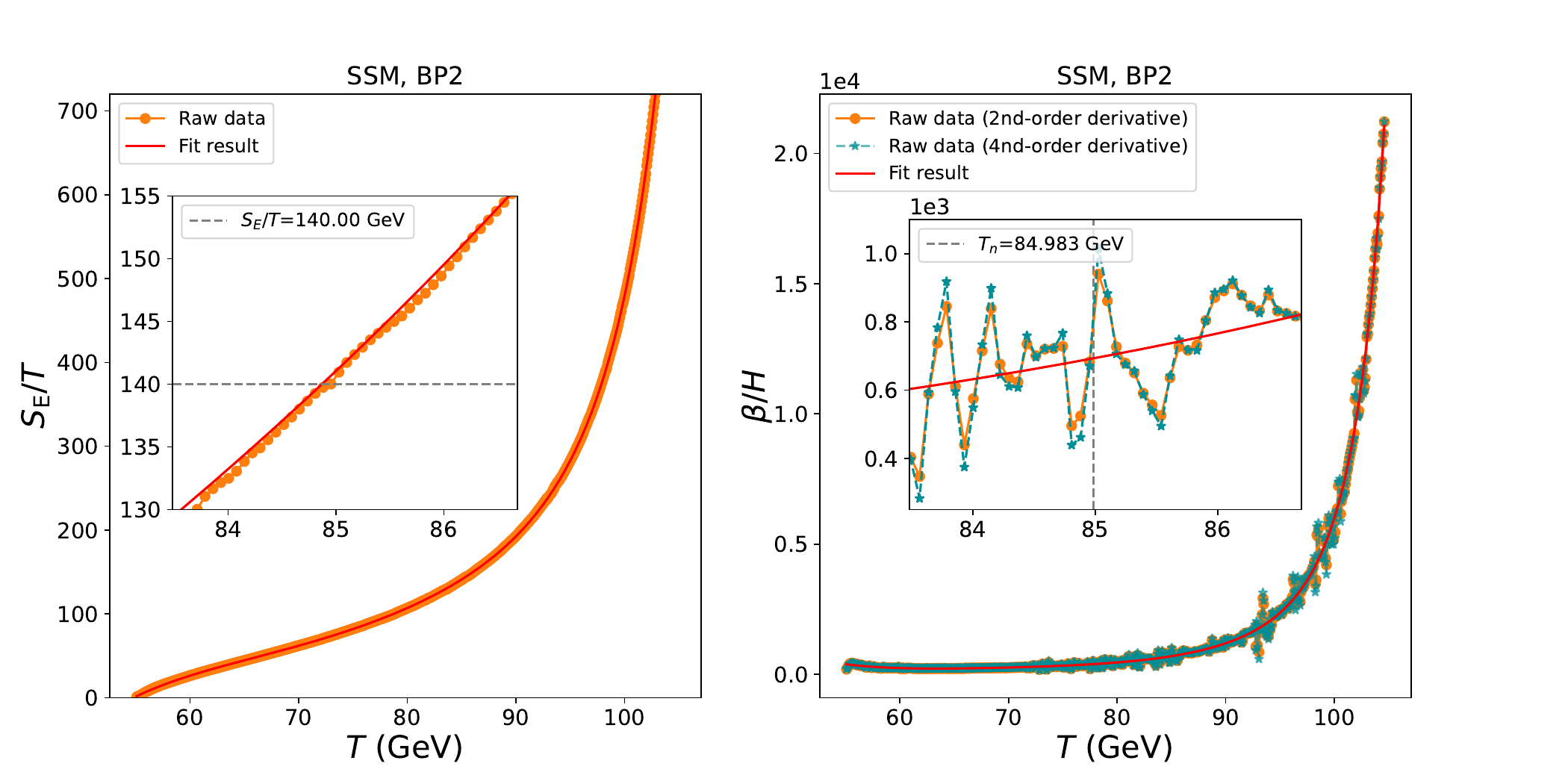}
    
    \caption{Top: The variation of $S_{E}/T$ and $\beta/H$ with temperature for BP1 in the SSM, where the dotted line represent the raw data calculated using \code{PhaseTracer2} and the $\beta/H$ values calculated from the raw $S_{E}/T$ data using ~\cref{eq: derivative}.
    Bottom: Same as top figure but for BP2. 
    The gray dashed lines represent the nucleation temperatures for each BP.
    }\label{fig: S/T}
\end{figure}

The numerical computation of the action inherently contains instability components, independent of the chosen algorithmic approach.
Fig.~10 of the \code{BubbleProfiler} manual~\cite{Athron:2019nbd} shows relative difference in $S_{E}$ between \code{BubbleProfiler} and \code{CosmoTransitions} and between \code{BubbleProfiler} and \code{AnyBubble}. The difference exhibits obvious oscillations as the effective potential varies. These results were obtained using the simplest one-dimensional potential, while for complex, multi-dimensional effective potentials, such numerical instabilities become more pronounced.
In \cref{fig: S/T}, we present the $S_{E}/T$ ratio as a function of $T$ for two benchmark points, BP1 and BP2, in the  SSM (which will be elaborated upon subsequently), which are calculated  utilizing \code{PhaseTracer2}. 
In principle, the action should undergo a smooth variation with changes in temperature, mirroring the continuous transformation of the effective potential as the temperature alters. While the action does appear to vary smoothly across the entire temperature range, upon closer inspection within a narrow interval, we observe a zigzag pattern indicative of numerical fluctuations in the sub-figures.

The prevalently used nucleation temperature $T_N$, at which the number of bubbles per Hubble volume attains unity, is defined  by 
\begin{equation} 
    \int_{T_N}^{\infty} \frac{{\rm d}T}{T}  \left( \frac{2\zeta M_{\rm PL}}{T} \right)^4 e^{-S_E/T} = \mathcal{O}(1),
\end{equation}
where $\zeta = \frac{1}{4\pi} \sqrt{45/(\pi g^*)}$ with $g^*$ being the effective number of relativistic degrees of freedom and $M_{\rm PL}$ is the reduced Planck mass~\cite{quiros1998finite}. Thus, $T_N$ can  be estimated from the function for $S_{E}/T$, 
\begin{equation} \label{eq: Tn}
    \frac{S_{E}(T_{N})}{T_{N}} \sim \mathcal{O}(140). 
\end{equation}
The nucleation temperatures for BP1 and BP2 are indicated by gray dashed lines in \cref{fig: S/T}. We see that the numerical fluctuations in $S_{E}$ can introduce sub-GeV errors in the determination of $T_N$. 

While this level of uncertainty may be considered acceptable, the situation deteriorates significantly when it comes to calculating the inverse phase transition duration time $\beta$, which is frequently normalized against the Hubble time at selected reference termpearture,
\begin{equation} \label{eq: beta}
    \frac{\beta}{H} = T\frac{{\rm d} (S_{E}/T)}{{\rm d} T}.
\end{equation}
Without knowing analytical expression of action, we can calculate derivative numerically using 
\begin{align} \label{eq: derivative}
f'(x) &\approx \frac{f(x+h) - f(x-h)}{2h}, ~\rm{(2nd}-\rm{order)}\\
f'(x) &\approx \frac{-f(x+2h) + 8f(x+h) - 8f(x-h) + f(x-2h)}{12h}, ~\rm{(4nd}-\rm{order)}
\end{align}
where $h$ represents a small step size. As $h$ approaches zero, the difference between $S_{E}(T+h)$ and $S_{E}(T)$ does not converge to zero, leading to instabilities and larger error. 

In the right planes of \cref{fig: S/T}, we present the $\beta/H$ for BP1 and BP2, where $h$ is set to 0.1 GeV. 
We can see that the uncertainties on $\beta/H$ in the SSM are substantial, which surpass 100\%, and sometime $\beta/H$ turns negative. Increasing the order of derivative calculation can not solve the problem. This is unacceptable, as $\beta/H$ is an important parameter in the calculation of gravitational wave caused by phase transition. 
For instance, the fitted formula for gravitational wave from (sw) and (turb) are given~\cite{Caprini:2009yp, PhysRevD.92.123009, Hindmarsh:2017gnf, PhysRevD.101.089902,Caprini:2019egz}
\begin{align} \label{eq: gw fitting}
    \Omega_{\rm sw}h^2 &= 2.65 \times 10^{-6} \left(\frac{H}{\beta}\right) \left(\frac{\kappa_v \alpha}{1 + \alpha}\right)\left(\frac{100}{g_{*}}\right)^{1/3} \left(\frac{f}{f_{\rm sw}}\right)^3 S_{\rm sw}(f_{\rm sw}) v_w\Upsilon(\tau_{sw}), \notag\\
    \Omega_{\rm turb}h^2 &= 3.35 \times 10^{-4} \left(\frac{H}{\beta}\right) \left(\frac{\kappa_{\rm turb} \alpha}{1 + \alpha}\right)^{3/2}\left(\frac{100}{g_{*}}\right)^{1/3} \left(\frac{f}{f_{\rm turb}}\right)^3 S_{\rm turb}(f_{\rm turb}) v_w, \notag\\
    f_{\rm sw} &= 1.9 \times 10^{-5} \frac{1}{v_w}\left(\frac{\beta}{H}\right) \left(\frac{T_{*}}{100 {\rm GeV}}\right) \left(\frac{g_{*}}{100}\right)^{1/6}~{\rm Hz}, \notag\\
    f_{\rm turb} &= 2.7 \times 10^{-5} \frac{1}{v_w}\left(\frac{\beta}{H}\right) \left(\frac{T_{*}}{100 {\rm GeV}}\right) \left(\frac{g_{*}}{100}\right)^{1/6}~{\rm Hz},
\end{align}
where $\alpha$ is the strength factor, $\kappa_{\rm sw}$, $\kappa_{\rm turb} \approx 0.1 \kappa_{\rm sw}$ denote the efficiency factor, $S_{\rm sw}$, $S_{\rm turb}$ represent spectral function and $\Upsilon(\tau_{\rm sw})$ is the suppression factor. 
The numerical errors in the calculation of $\beta$ could ultimately impact the resulting gravitational wave spectrum. 

Apart from numerical instability, the computation time poses an additional challenge in numerical integration operations that involve the action. As highlighted in Ref. \cite{Cai:2017tmh, Athron:2022mmm,Athron:2023rfq}, the nucleation temperature may not serve as the most optimal physical quantity for characterizing the phase transition. Instead, the percolation temperature $T_P$, defined as the temperature at which the fraction of false vacuum decreases to 70\%, offers a more accurate reflection of the actual scenario:
\begin{align} \label{eq: h(t)}
  h(T_P)  &= {\rm{exp}}\left[-\int^{T_P} _{T_C} \Gamma(t')V(t',t)\frac{{\rm d}t'}{{\rm d}T'} {\rm d} T'\right] = 0.7, 
\end{align}
where $V(t',t) = \frac{4\pi}{3} [\int ^t _{t'} v_{\rm w}(\tau) d\tau]^3$, $v_{\rm w}$ is the bubble wall velocity and $\Gamma$ is the nucleation rate per volume per time
\begin{equation} \label{eq:nucleation rate}
    \Gamma \sim e^{-S_{E}/T}.
\end{equation}
Calculating $T_P$ using the bisection method necessitates solving the aforementioned integral numerous times, consuming a considerable amount of time. A single integral may require hundreds of action calculations, and the bisection method demands dozens of such integrals.
 
By and large, the calculations of transition parameters related to the action encounters two challenges: strong oscillations and excessive computation time.
Significant efforts have been dedicated to improving the accuracy of action calculations~\cite{Akula:2016gpl,Andreassen:2016cff,Brown:2017cca,Espinosa:2018szu,Espinosa:2018hue,Braden:2018tky,Piscopo:2019txs,Sato:2019axv,Bardsley:2021lmq}, but refining accuracy often comes at the cost of either a narrower range of applicability or an increase in computational time.
Fitting the action function of temperature appears to be an alternative solution to address these problems. In \code{PhaseTracer2}, a linear fit is employed in the vicinity of the reference temperature for calculating $\beta$. In \code{Vevacious}~\cite{Camargo-Molina:2014pwa}, to estimate the upper bound on the survival probability, it uses
$(T_C -T)^2$ times a polynomial in $T$ to approximate the $S_E^{\rm straight}$,  which represents the bounce action
along a straight path in field space from the false vacuum to the true vacuum. 
Recently, \code{PT2GWFinder}~\cite{Brdar:2025gyo} released their code, which employs a same fitting scheme near the $T_C$. Here we utilize the fitting function across the entire temperature range and demonstrate its effectiveness in both the high temperature  approximated potential and a more complex scenario. We provide a detailed analysis of the advantages offered by the action fit. Additionally, we propose a method that uses neural networks to predict the action function directly from model parameters.


\section{Fitting of the action function} \label{sec: fitting}

There has been continuous effort developing semi-analytical expression for the action. To date, only one-dimensional 4-order polynomial effective potentials have yielded such semi-analytical formulations, which demonstrate excellent agreement with numerical computations~\cite{Dine:1992wr,Adams:1993zs,Gouttenoire:2022gwi,Matteini:2024xvg}. 
For instance, the action of 
\begin{equation} 
    V(\phi) = a \phi^2 - b\phi^3 + \frac{\lambda}{4} \phi^4
\end{equation}
is given by
\begin{align}
    &S_E = \frac{13.72 a^{3/2}}{b^2} f(\delta/2),~~~~~\delta=\frac{2\lambda a}{b^2}\\
    &f(x) = 1+ \frac{x}{4}\left[1+\frac{2.4}{1-x} + \frac{0.26}{((1-x)^2} \right].
\end{align} 
The analytical expression enables precise and stable calculation of the action, thereby circumventing the numerical instabilities discussed in the previous section. However, no semi-analytical solutions currently exist for more general classes of effective potentials.

The action as a function of temperature diverges when $T\to T_C$, and it may be easier to perform function fit after eliminating this divergence. Therefore, we first need to extract the divergent behavior of the action curve. At the critical temperature $T_C$, the two minimums degenerates, so this divergent trend can be described by the thin-wall approximation of the action. 

In the thin-wall approximation, where the difference in effective potential between the false vacuum $\phi_f$ and the true vacuum $\phi_t$, defined as $\epsilon = V_{\rm eff}(\phi_f; T) -  V_{\rm eff}(\phi_t; T)$, is much smaller than the height of the barrier, the action can be expressed in an analytical form. Under this assumption, Eq. (\ref{eq: bubble profile}) can neglect the viscous damping term, simplifying to:
\begin{equation} 
    \frac{{\rm d}^2 \phi}{{\rm d} r^2}  = \frac{\partial V_{\rm eff}(\phi;T)}{\partial \phi},
\end{equation}
which gives 
\begin{equation}
    \frac{d \phi}{d r}  = {\sqrt{2 V_{\rm eff}(\phi; T)}}.
\end{equation}
By defining the surface tension of the bubble
\begin{equation}
    \sigma = \int_0 ^\infty dr \left[\frac{1}{2}\left(\frac{d \phi}{d r}\right)^2 + V_{\rm eff}(\phi; T)\right] = \int_{\phi_t}^{\phi_f}d\phi \sqrt{2V_{\rm eff}(\phi; T)} ,
\end{equation}
the Euclidean action can be expressed as 
\begin{align}
        S_{E} &=  4 \pi \int^{+\infty}_{0}r^2{\rm d}r~\left[\frac{1}{2}\left(\frac{\partial \phi}{\partial r}\right)^{2} + V_{\rm eff}(\phi;T)\right], \notag \\
        &=4 \pi R^2 \sigma - \frac{4}{3} \pi R^3 \epsilon.
\end{align}
where $R$ is the radius of the critical bubble and can be calculated by minimization of $S_{E}$,
\begin{equation}
    R = \frac{2 \sigma}{\epsilon}.
\end{equation}
Collecting all the expressions together, $S_{E}$ can be reorganized as 
\begin{equation} \label{eq: S_E fitting expression}
    S_E = \frac{16 \pi \sigma^3}{3 \epsilon^2}.
\end{equation}
When the temperature deviates slightly from the critical temperature, $\epsilon$ is close to zero and significantly lower than the barrier height. We can perform a Taylor expansion in the vicinity of this temperature,
\begin{align}
    \epsilon &= V_{\rm eff}(\phi_f;T) - V_{\rm eff}(\phi_t;T), \notag \\
    &=\left( \left.\frac{\partial V_{\rm eff}(\phi_f;T)}{\partial T}\right|_{T = T_C} - \left.\frac{\partial V_{\rm eff}(\phi_t;T)}{\partial T}\right|_{T = T_C}\right) \left(T - T_C\right),
\end{align}
where we have utilized the fact that the vacuum corresponds to the minimum of the effective potential. Substituting this into Eq. (\ref{eq: S_E fitting expression}), we can derive the following heuristic form:
\begin{equation}
    S_E \sim \frac{f(T)}{(T-T_C)^2},
\end{equation}
where $f(T)$ is closely related to the specific form of the effective potential, such as shown in the semi-analytical expression of one dimensional model obtained in the $T_C$ limit in~\cite{Enqvist:1991xw}. It is challenging to derive $f(T)$ analytically for an arbitrary model, so we employ the polynomial fitting formula
\begin{equation} \label{eq:fit_fun}
    S_E = \frac{1}{(T-T_C)^2}\displaystyle\sum_{i=0}^{n_{\rm order}} q_i T^i
\end{equation}
to numerically obtain a reliable estimate for the Euclidean action. In this case, one can get the expression for the inverse phase transition duration time,
\begin{equation} 
    \frac{\beta}{H} = 
\frac{
  1}{ T (T - T_C)^3} \left[ \displaystyle \sum_{i=1}^{n_{\mathrm{order}}} q_i i T^{i} (T - T_C)   - \sum_{i=0}^{n_{\mathrm{order}}} q_i T^i (3T - T_C) \right], 
\end{equation}
and mitigate significant computational errors.

Note that while the factor $1/(T-T_C)^2$ is derived specifically for the thin-wall approximation, the fitting of \cref{eq:fit_fun} is performed across the entire temperature range - spanning both thin-wall and thick-wall regimes. In the thick-wall limit where $T$ is far from $T_C$, this factor converges to a finite value and therefore has no impact on the fit. Consequently, the resulting fit remains valid throughout the complete temperature range.

We have implemented this polynomial fitting method in the latest version of \code{PhaseTracer2}, and a detailed guide on its usage is provided in Appendix A. Appendix B provides instructions for applying it within \code{CosmoTransitions}.

\subsection{One-dimensional model }

In order to validate the polynomial fitting approach, we utilize the high temperature  approximated potential in which the action can be accurately computed using the shooting method, thereby ensuring high accuracy. The effective potential is given by 
\begin{align}
    V_{\rm eff}(\phi;T) = (c T^2 - m^2) \phi^2 + \kappa \phi^3 + \lambda \phi^4
\end{align}
where $c$, $m$, $\kappa$ and $\lambda$ are model parameters. It is representative of models where the potential barrier arises from renormalizable tree-level interactions between the Higgs and new scalar fields~\cite{Chung:2012vg}.

\begin{figure}[thbp!]
    \centering
    \includegraphics[width=0.9\textwidth]{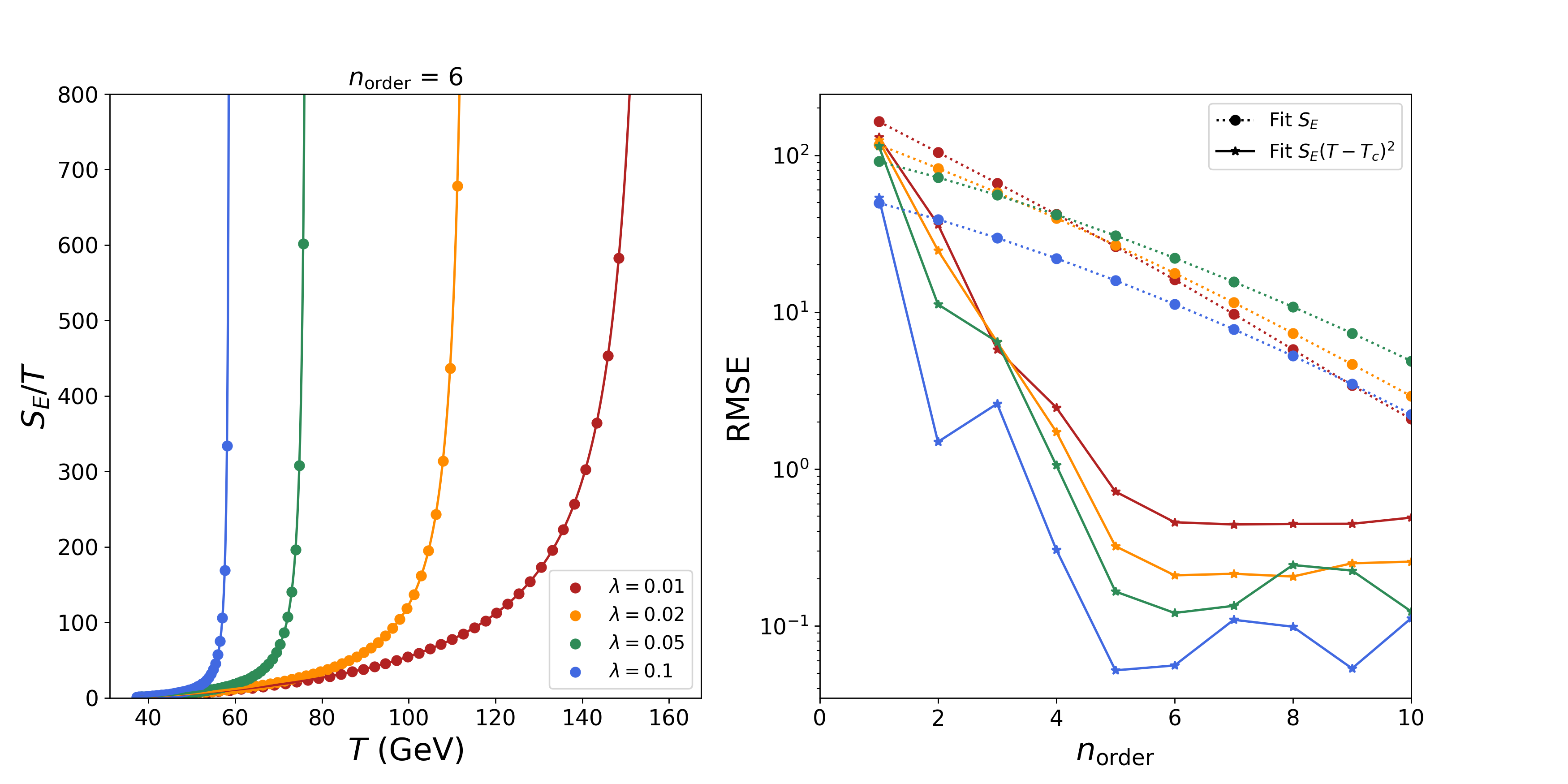}
    \includegraphics[width=0.9\textwidth]{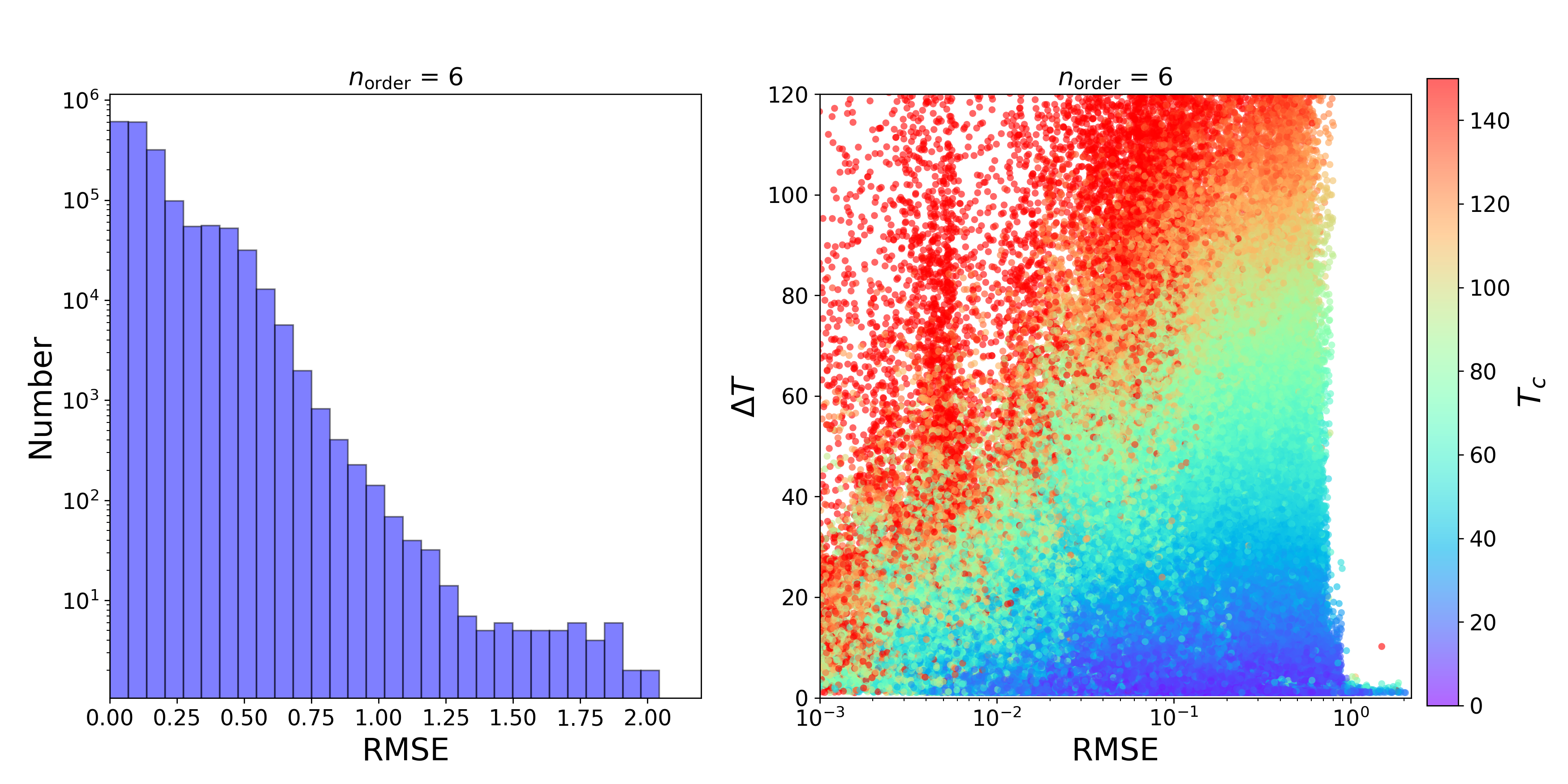}
    \caption{Fitting results for the one-dimensional high temperature  approximated potential. Top left: $S/T$ as a functions of $T$ for different benchmark points. The dots represent raw data from the shooting method, while the curves show the fitted results. Top right: the RMSE for different fitting orders. The dotted lines correspond to direct fits of $S_E$ directly, while the solid lines represent fits to $S_E(T-T_C)^2$.  Bottom left: the distribution of the RMSE from a random parameter scan. Bottom right: the correlation between the RMSE and the temperature interval $\Delta T$.  }\label{fig: 1D fiiting results}
\end{figure}

The performance of the fitting are presented in \cref{fig: 1D fiiting results}. In the top left panel, we exhibit the raw results of the action calculation for various benchmark points with $m^2=100,~\kappa=-10,~c=0.1$ and different values of different $\lambda$. 
The fitting results using \cref{eq:fit_fun} with $n_{\rm order}=6$ are depicted by lines, which align very well with the raw data. 
To quantify the degree of agreement, we utilize the root mean square error (RMSE), which is given by:
\begin{align}
    {\rm RMSE} = \sqrt{ \frac{1}{n} \sum_{i=1}^{n} \left(\frac{S_E(T_i)}{T_i} - \frac{\hat{S}_E(T_i)}{T_i}\right)^2 }
\end{align}
where $S_E(T_i)$ represents the action calculated directly using the numerical tool at $T_i$, while $\hat{S}_E(T_i)$ denotes the action obtained from the fitting results at the same temperature.
The top right panel displays 
the RMSE of various $n_{\rm order}$ values for the same benchmark points. The results directly fitting $S_E(T_i)$, instead of using \cref{eq:fit_fun}, are also shown. As expected, the RMSE drops with the increasing of $n_{\rm order}$. Without the factor $(T-T_C)^2$, the rate of decline in RMSE is slower, reaching values greater than 10 for $n_{\rm order}=10$. Conversely, using \cref{eq:fit_fun}, the RMSE initially decreases steeply and then flattens out for $n_{\rm order}>5$. It suggests that the a sixth-order polynomial fitting incorporating the factor $(T-T_C)^2$ provides an excellent description of the action function.

In the bottom left panels of \cref{fig: 1D fiiting results}, we present the distribution of the RMSE from a random scan in following parameter space,
\begin{align}
    c \in [0,2],~ m^2 \in [0,200],~\lambda \in [0,2],~\kappa \in [-30,0].
\end{align}
It shows that the majority of samples (99.987\%) exhibit an RMSE below 1, with a maximum value of 4.2.  These values are negligible compared to the typical magnitude of $S/T\sim 140$ in our region of interest, as seen in \cref{eq: Tn}. 
The bottom right panel displays the samples on the RMSE-$\Delta T$ plane, with color coding indicating $T_C$. Here, the temperature interval is defined as
\begin{align}
    \Delta T = T_C - T_{\rm min},
\end{align}
where $T_{\rm min}$ denotes the minimum temperature within the overlapping range of the two coexisting phases. The action can only be computed within this interval. The RMSE exceeds 1 only for $\Delta T <2$ GeV,  resulting from insufficient sampling resolution in this narrow temperature range. However, such small temperature intervals correspond to physically uninteresting fast transitions, they consistently exhibit low $T_C$ values.

\begin{table}[!th]
\centering
\begin{tabular}{clccccccccc}
\hline\hline
                     & & $c$                  & $m^2$                & $\kappa$             & $\lambda$             & $T_C$ & $T_N$ & $T_P$  & Time \\
\hline
\multirow{2}{*}{\textbf{BP1}} & without action fit  & \multirow{2}{*}{0.1} & \multirow{2}{*}{100} & \multirow{2}{*}{-10} & \multirow{2}{*}{0.01} & 161.2     & 125.7     & 122.0            & 5.60s    \\
                     & with action fit &                      &                      &                      &                       & 161.2     & 125.7     & 122.0     &  0.04s   \\
\hline
\multirow{2}{*}{\textbf{BP2}} & without action fit  & \multirow{2}{*}{0.1} & \multirow{2}{*}{100} & \multirow{2}{*}{-10} & \multirow{2}{*}{0.02} & 116.2     & 101.5     & 99.6     &  7.54s    \\
                     & with action fit &                      &                      &                      &                       & 116.2     & 101.4     & 99.6     & 0.05s   \\
\hline
\multirow{2}{*}{\textbf{BP3}} & without action fit  & \multirow{2}{*}{0.1} & \multirow{2}{*}{100} & \multirow{2}{*}{-10} & \multirow{2}{*}{0.05} & 77.5     & 73.0     & 72.4       & 10.8s    \\
                     & with action fit &                      &                      &                      &                       & 77.5     & 73.0     & 72.4     & 0.05s   \\
\hline
\multirow{2}{*}{\textbf{BP4}} & without action fit & \multirow{2}{*}{0.1} & \multirow{2}{*}{100} & \multirow{2}{*}{-10} & \multirow{2}{*}{0.1} & 59.2     & 57.4     & 57.2           & 11.46s    \\
                     & with action fit &                      &                      &                      &                       & 59.2     & 57.4     & 57.2       &   0.05s  \\
\hline\hline
\end{tabular}
\caption{The input parameters and transition parameters calculated with and without action fit for the high temperature  approximated potential. The last column shows the corresponding computation time, measured on a MacBook Pro with an Intel Core i5 processor.}
\label{tab:1d_BP}
\end{table}

In \cref{tab:1d_BP}, we compare the computational time to calculate the phase transition temperature with and without the action fit method, for the four benchmark points presented in \cref{fig: 1D fiiting results}. The $T_P$ calculation \cref{eq: h(t)} involves an integral that requires many action evaluations, which is computationally intensive. Using an action fit reduces this cost by approximately two orders of magnitude, cutting the computation time from several seconds to a fraction of a second.

In summary, through polynomial fitting of \cref{eq:fit_fun}, we can accurately describe the functional dependence of $S/T$ on T, which subsequently enables the calculation of other phase transition parameters accurately and quickly.

\subsection{Two-dimensional model }

Now we turn to a physical model, SSM, which is wildly used in instructional studies of phase transition. The effective potential, a function of the Higgs vacuum expectation value 
$h$ and the vacuum expectation value of the additional scalar field $s$, contains serial parts: 
\begin{equation} \label{eq: V_eff}
V_{\rm eff}(h, s; T) = V_{0}(h, s) + V_{\rm CW}(h, s) + V_{\rm CT}(h, s) +  V_{\rm 1T}(h, s; T) + V_{\rm ring}(h, s; T), 
\end{equation}
where $V_0$, $V_{\rm CW}$, $V_{\rm CT}$, $V_{\rm 1T}$ and $V_{\rm ring}$ are the tree level potential, the one-loop Coleman-Weinberg correction, the corresponding counter term, the one-loop thermal correction and the resummed daisy correction, respectively. The tree level potential can be written as 
\begin{equation}\label{eq:xsm_tree}
     V_{0}(h,s) = -\frac{\mu_{H}^{2}}{2}h^2 + \frac{\lambda_{H}}{4}h^4 - \frac{\mu_{S}^2}{2}s^2 + \frac{\lambda_{S}}{4}s^4 + \frac{\lambda_{HS}}{4}h^2s^2.
\end{equation}
The parameters $\mu_{H}$ and $\lambda_{H}$ can fixed by the tadpole condition and so there remains three free parameters. By the tree level mass relationship, we could choose the three input parameters as $m_{S}$, $\lambda_{S}$, and $\lambda_{HS}$. For the  one-loop zero-temperature correction, we choose the OS-like scheme and the Landau gauge~\cite{Quiros:1999jp} and it takes a form
\begin{equation} \label{eq: V_CW}
\begin{aligned}
V_{1}(h, s)  =& V_{\rm CW}(h, s) + V_{\rm CT}(h, s) \\
 =& \sum_{i} (-1)^{s_{i}} \frac{g_{i}}{64 \pi^2}  \left\{ m_{i}^4(h, s)\left[\log \frac{m_{i}^2(h, s)}{m_{i}^2(v_{h}, v_{s})} -\frac{3}{2} \right] + 2m_{i}^2(h, s)m_{i}^2(v_{h}, v_{s}) \right\},
\end{aligned}
\end{equation}
where $i$ represent the Higgs boson, additional singlet scalar field, top quark, photon, W and Z bosons, $s_i$($g_i$) is the spin (number of degrees of freedom) of the corresponding particle. We have neglected the contributions of light fermions and Goldstone bosons. The former is due to its relatively small coupling constant with the scalar field, while the latter is due to $\left.m_{G^{\pm,0}}\right|_{h=v_h,s=0}=0$ and $\left. \partial m_{G^{\pm,0}}/\partial h \right|_{h=v_h,s=0}\neq0$, leading the logarithmic divergent of the 
second derivative of $V_{\rm CW}$ at VEV of zero-temperature. The impact of fixing the Goldstone catastrophe, as well as choosing other renormalization scheme, can be found in Ref.~\cite{Braathen:2016cqe, Athron:2022jyi}.

The one-loop finite temperature correction is deduced from the finite-temperature field theory
\begin{equation} \label{eq: one-loop effective potential at finite temperature}
    V_{\rm 1T}(h, s) = \frac{T^{4}}{2 \pi^2} \left[\sum_{B} g_{B}J_{B}\left(\frac{m_{B}(h, s)}{T}\right) + \sum_{F} g_{F}J_{F}\left(\frac{m_{F}(h, s)}{T}\right)\right],
\end{equation}
where $J_{B}$, $J_{F}$ are the relevant thermal distribution functions for the bosonic and fermionic contributions, respectively~\cite{Quiros:1999jp}. Because of the zero-mode contribution, the perturbation method is broken~\cite{senaha2020symmetry, Curtin:2016urg} and so the multi-loop contributions must be resummed. In this paper, the resummed method proposed by Parwani is adopted through replacing the tree-level masses $m_{i}^2$ with the thermal masses $m_{i}^2(T) = m_{i}^2 + d_{i}T^2$~\cite{Parwani:1991gq}, where $d_{i}T^2$ is the contribution of the one-loop thermal mass and can be found in \cite{Basler:2018cwe}.

The high temperature approximation of this model has the form
\begin{equation} 
    V_{\rm HT}(h, s) = -\frac{\mu_{H}^{2}+c_H T^2}{2}h^2 + \frac{\lambda_{H}}{4}h^4 - \frac{\mu_{S}^2+c_S T^2}{2}s^2 + \frac{\lambda_{S}}{4}s^4 + \frac{\lambda_{HS}}{4}h^2s^2,
\end{equation}
where $c_H$ and $c_S$ include the thermal contribution. The potential barrier also  arises from renormalizable tree-level interactions.
In the following, we use the full effective potential in \cref{eq: V_eff} instead of the approximation to demonstrate the accuracy and universality of the action fit method.

\begin{figure}[htbp!]
    \centering
    \includegraphics[width=0.9\textwidth]{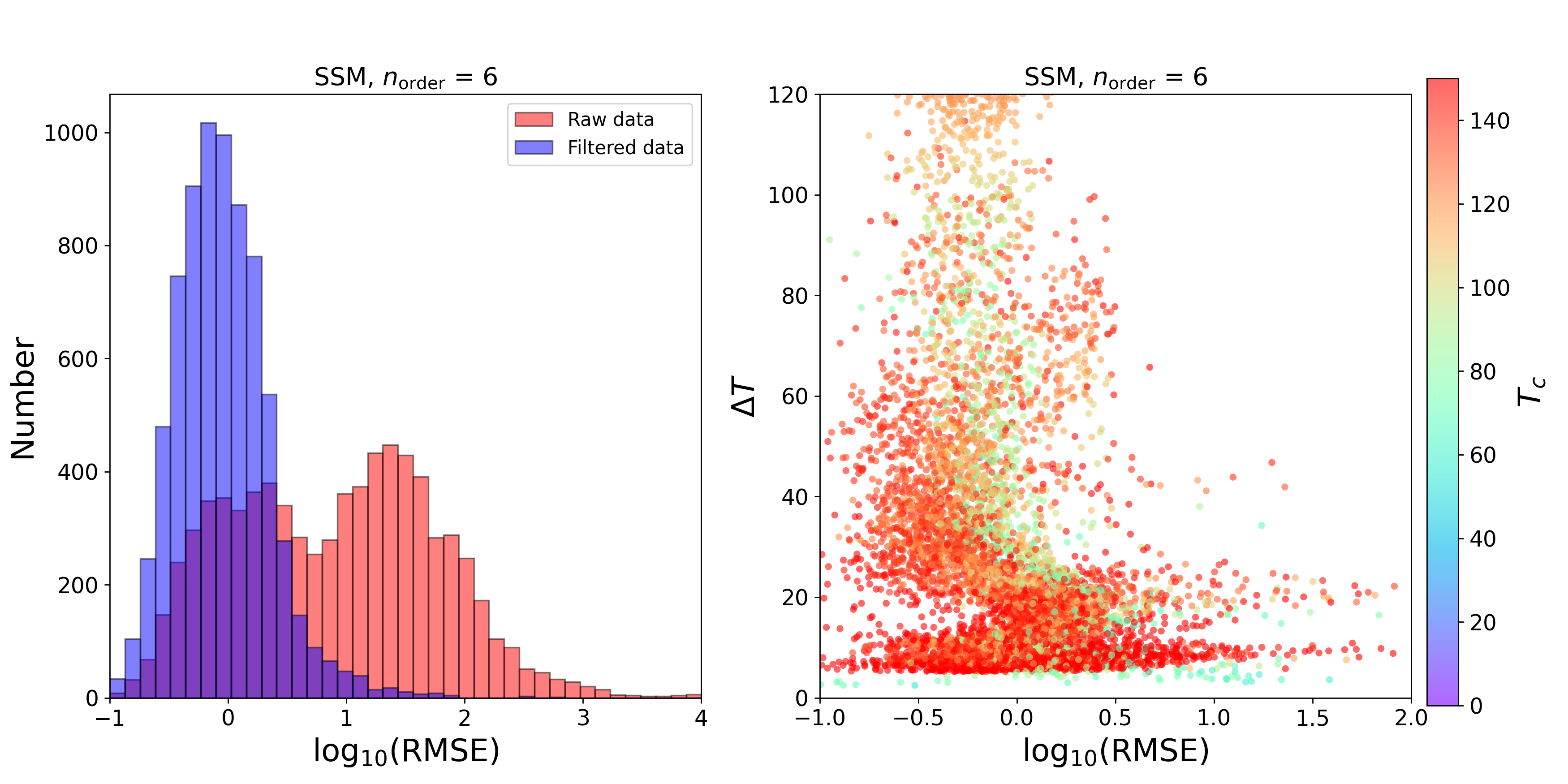}
    \caption{Fitting results for the SSM model. The left panel displays the distribution of the RMSE from a random parameter scan. The blue and red histograms represent the distributions before and after data trimming, respectively. The right panel illustrates the relationship between RMSE and the temperature range $\Delta T$, with color indicating the Transition $T_C$. }\label{fig: 2D fiiting results}
\end{figure}

\begin{figure}[htbp!]
    \centering
    \includegraphics[width=0.9\textwidth]{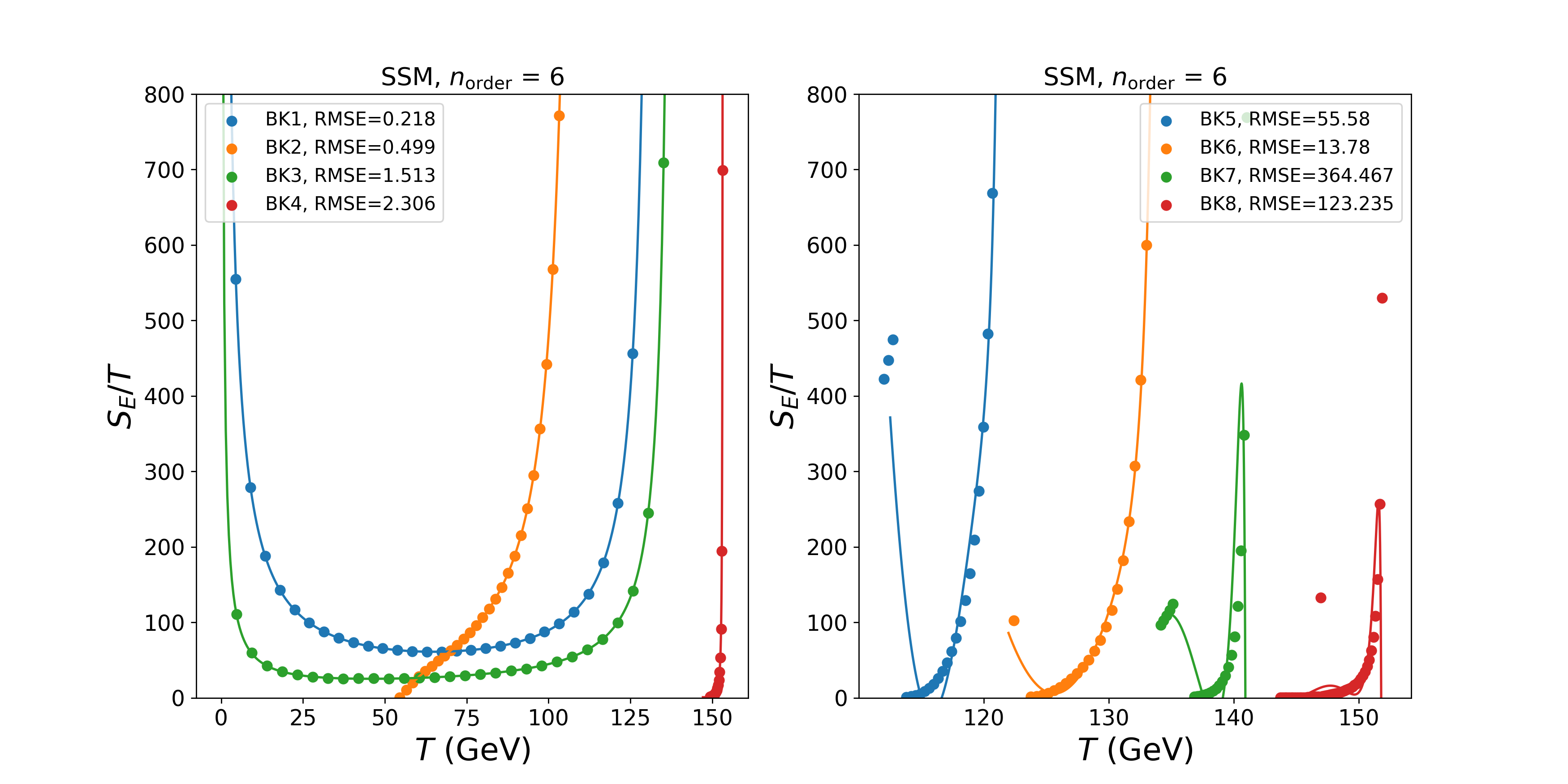}
    \caption{The action versus temperature for benchmark points in the SSM model. The dotted points represent raw data obtained from PhaseTracer, and the curves correspond to the fitted functions. The left panel displays cases with small RMSE, while the right panel shows those with large RMSE.}\label{fig: 2D BKs}
\end{figure}

We perform a random scan in the following parameter space, 
\begin{equation} 
10 \gev < m_S < 120 \gev,~~ 0.01<\lambda_S <0.2,~~0.15<\lambda_{HS}< 0.6.
\end{equation}
The fitting results are presented in \cref{fig: 2D fiiting results}.
The left panel displays the distribution of the RMSE across the sampled points, where the blue histogram represents the results from direct fitting without applying node trimming to the action curve. It can be observed that approximately 47.8\% of the points exhibit an RMSE greater than 10, a situation much worse than that of the high temperature  approximated potential. This discrepancy arises because the calculation of the action for a complex high-dimensional potential is not always numerically stable. 
The left panel of \cref{fig: 2D BKs} displays several benchmark points (listed in \cref{SSM BP}) with low RMSE values, all of which exhibit satisfactory behavior. In contrast, the bottom right panel presents cases with high RMSE, where abnormal points in the action data are clearly visible, adversely affecting the quality of the fit. 
The anomalous data points primarily occur at the left end of the action curve, corresponding to the temperature at which one of the minima vanishes. This makes it prone to inaccurate action values. Occasionally, such anomalies also appear within the mid-range of the temperature interval.

Therefore, it is necessary to trim the data prior to performing the fit. After various attempts, we find that the following criterion effectively removes abnormal points:
\begin{equation} \label{eq: selection } 
S(T_i) - S(T_{i-1}) < 0 \quad \text{and} \quad |S(T_i) - S(T_{i-1})| < k \cdot |S(T_{i-1}) - S(T_{i-2})|
\end{equation}
where $T_i > T_{i+1}$, the temperature intervals are fixed, and $k$ is manually set to 2. This criterion is based on the expectation that, as temperature decreases from $T_C$ to $T_{\rm min}$ (the minimum temperature of the overlapping range of the two phases), the action should decrease monotonically, and the rate of decrease should slow. In extreme cases where the action values are computed correctly but do not satisfy this criterion, the calculation can fall back to the traditional method.

The red histogram in the top left panel of \cref{fig: 2D fiiting results} shows the RMSE distribution after applying this filtering criterion. It can be seen that approximately 94.83\% of the samples now exhibit an RMSE below 5, making them suitable for further analysis. The error here remains larger than that in the one-dimensional high temperature  approximated potential, primarily due to the inherent and unavoidable inaccuracies in calculating the action for complex potentials. This outcome is sufficient to achieve the dual goals of reducing computational time and improving accuracy.

\begin{table}
\centering
	\begin{tabular}{ccccccc}
		\hline
        \hline 
		& $\lambda_{S}$ & $\lambda_{HS}$ & $m_{S}$ (GeV)  & $T_C$ (GeV) & $T_N$ (GeV) & $\beta$\\
		\hline
        
		$\textbf{BP1}$ &~~0.127918~~     & ~~0.590136~~        & ~~117.725 ~~  & ~~134.694~~ & ~~112.598~~ & ~~798.302~~  \\  
		$\textbf{BP2}$ &~~0.190164~~     & ~~0.450307~~        & ~~84.384~~   & 112.994  & 84.983 & 686.925   \\  
		$\textbf{BP3}$ &~~0.108695~~     & ~~0.548401~~        & ~~115.245~~   & 139.782 & 125.689 & 1596.97   \\  
		$\textbf{BP4}$ &~~0.0929279~~     & ~~0.171384~~        & ~~62.2088~~   & 153.371 & 152.902 &108555.0   \\  
		\hline
        \hline 
	\end{tabular}
\caption{The input parameters and transition parameters for the benchmark points in the SSM. }
\label{SSM BP}
\end{table}

In \cref{tab:data}, we compare the values of $\beta$ obtained with and without the action fitting method, using different $h$ in \cref{eq: derivative}. Each approach is executed 10 times, and the mean value along with the standard deviation are computed. Although the nucleation temperature $T_N$ varies slightly across runs, we fix $T_N = 109.27~\gev$ for consistent comparison of the results.
The results indicate that with a large step size $h$, the calculated $\beta$ is stable but statistically deviates from the central value. In contrast, a small step size $h$ leads to increased numerical instability. On the other hand, using the action fitting method yields both accurate and stable results for $\beta$.

\begin{table}[htbp]
  \centering
  \begin{tabular}{cccccc}
  \hline
  \hline
    & $h = 1$ & $h = 0.1$ & $h = 0.01$ & $h = 0.001$ & Fitting \\
    \hline
1	&	1611.63	&	1544.54	&	1645.27	&	1556.33	&	1601.22	\\
2	&	1610.61	&	1596.06	&	1624.83	&	1622.45	&	1589.33	\\
3	&	1624.05	&	1600.04	&	1624.43	&	1605.21	&	1595.67	\\
4	&	1612.21	&	1593.78	&	1531.95	&	1949.19	&	1596.19	\\
5	&	1623.95	&	1600.50	&	1533.05	&	1553.57	&	1596.14	\\
6	&	1611.03	&	1543.98	&	1645.10	&	1624.50	&	1595.06	\\
7	&	1624.00	&	1593.41	&	1532.45	&	1618.53	&	1595.50	\\
8	&	1610.52	&	1594.68	&	1533.93	&	1610.89	&	1594.97	\\
9	&	1612.21	&	1593.78	&	1644.55	&	2089.34	&	1595.95	\\
10	&	1624.04	&	1600.04	&	1645.29	&	1557.93	&	1596.20	\\
\hline
Mean	&	1616.43$\pm$6.55	&	1586.08$\pm$22.2	&	1596.09$\pm$54.99	&	1678.79$\pm$184.61	&	1595.62$\pm$2.84	\\
    \hline
  \end{tabular}
  \caption{Comparison of $\beta$ values computed using different step sizes $h$ and the action fit method for BP3. The mean and standard deviation (over 10 runs) are listed in the bottom row.}
  \label{tab:data}
\end{table}

\section{Neural network fitting of the action function}

A further advantage of employing action fitting is its compatibility with machine learning frameworks. Previous attempts have been made to apply machine learning methods to improve the investigation of cosmological phase transition~\cite{Piscopo:2019txs,Jinno:2018dek,Searle:2025cnj,Balassa:2025gwh}. 
Ref.~\cite{Piscopo:2019txs} employs neural networks to solve differential equations relevant to bounce action computations. Ref.~\cite{Jinno:2018dek} trains a neural network to predict the bounce action from one-dimensional potentials, which takes 15 nodes of the potential as input. With action fitting, we can establish neural networks from model parameters to the action curve, rather than computing individual action values point-by-point. Moreover, the learned result can be efficiently validated by testing selected points along the predicted curve, which is straightforward and computationally inexpensive.

We employ the conventional fully connected neural network as well as the Kolmogorov-Arnold Network~(KAN)~\cite{Liu:2024swq} to predict the action curve.
The entire learning process is divided into three main steps:
\begin{itemize}
    \item First, it is essential to distinguish parameter regions where valid first-order phase transition occurs. In the  one-dimensional high temperature  approximated potential or the SSM, a portion of the parameter space does not exhibit a first-order phase transition, or the temperature interval $\Delta T$ between two overlapping phases is extremely narrow. Studying the EWPT in such regions is meaningless, and obtaining a reliable action fit is impossible. Therefore, the first step is to identify these irrelevant parameter regions.
    \item Next, we need to predict the overlap temperature range of the two phases, i.e. $T_C$ and $T_{\rm min}$. These two parameters are crucial in addition to the fitting coefficients, as they define the valid temperature range of the fitted function. Moreover,  $T_C$  will explicitly appear in the final fitting formula.
    \item Finally, the coefficients of the polynomial are predicted. Due to the significant difference in the orders of magnitude of the coefficients being fitted, we chose to utilize the neural network to reconstruct all the coefficients piece by piece. Specifically, we first train the network to determine the signs of the coefficients, then train it to predict the direction of the vector representing the action polynomial in a Hilbert space, and finally train it to predict the magnitude of that vector. Once the polynomial coefficients are obtained, the complete action function is reconstructed. Special points, such as the nucleation temperature $T_N$ , can then be selected to validate the accuracy of the machine learning predictions. 
\end{itemize}

\subsection{Classification of the regions of phase transition}

Here we utilize the  one-dimensional high temperature  approximated potential to explore how machine learning can capture features of the action curve. As described in Sec. 3.1, the model features four input parameters: $c$, $m$, $\kappa$ and $\lambda$. We train the neural network using randomly sampled parameter points.

The identification of parameter regions exhibiting first-order phase transitions is straightforward, given that these regions are continuous and clearly delineated. 
We use a fully connected neural network with four input neurons~(the inputs are $c$, $m$, $\kappa$ and $\lambda$), three hidden layers with $32$ neurons, and an output layer with two neurons.
The two neurons in the output layer serve as the activation units for the binary classification problem.
The parametric rectified linear unit~(PReLU) function~\cite{He:2015dtg} is used as the activation function between each layers.
For the output layer, a `softmax' function is applied.
The loss function is chosen as the cross entropy.
Throughout this paper, neural networks are implemented using the \code{PyTorch} package~\cite{Paszke:2019xhz}.

The training set contains $512,000$ samples, and the validation set contains $130,320$ samples.
The z-score standardization~\cite{10.1214/009053604000000265} is used for the datasets.
Taking the input $c$ as an example, in the dataset, $\hat{c}$ is used instead of $c$ where $\hat{c}=\left(c - \bar{c}\right)/\sigma_c$, where $\bar{c}$ and $\sigma_c$ are the average and standard deviation of $c$ over the dataset.

In this work, we employ early stopping to prevent overfitting. 
Specifically, we monitor the validation dataset and terminate the training process when the loss function ceases to show significant improvement. 
The model checkpoint with minimal validation loss is selected as the final training outcome.
In this subsection, the model is trained for $1000$ epochs.

The final classification results show that 228 samples of 130,320 validation samples are misclassified, resulting in an accuracy of 99.8\%. 
The misclassified samples are mainly located at the boundary of two categories, making accurate classification difficult. 
The corresponding phase transition strength is sufficiently weak, thus having a negligible impact on the study of gravitational waves or matter-antimatter asymmetry.

\subsection{Fitting of \texorpdfstring{$T_C$}{Tc} and \texorpdfstring{$T_{\rm min}$}{Tmin}}

\begin{figure}[thbp!]
\centering
\includegraphics[width=0.6\hsize]{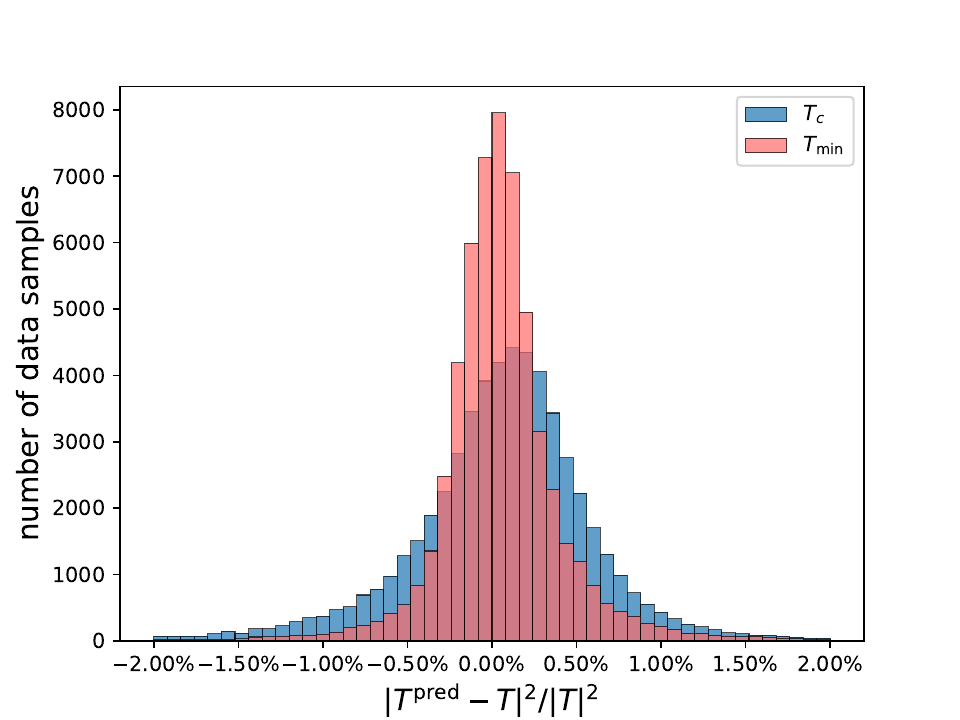}
\caption{Distribution of prediction errors for the critical temperature $T_C$ (blue) and the minimum temperature $T_{\rm min}$ (orange).
}\label{fig:TCErrorHist}
\end{figure}

The predictions of $T_C$  and  $T_{\rm min}$  are also relatively direct, as both temperatures vary continuously within certain bounds. We use a fully connected neural network with four input neurons~(the inputs are $c$, $m$, $\kappa$ and $\lambda$), three hidden layers with $32$ neurons, and an output layer with two neurons~(the two neurons are the predictions of $T_C$ and $T_{\rm min}$).
The PReLU function is used as the activation function between each layers.
The loss function is the mean absolute error~(mae) function.
Since only parameters within the phase transition region are used for training, the dataset is smaller than the one used in the previous subsection.
The training set contains $512,000$ samples, and the validation set contains $56,711$ samples.
In this subsection, the model is trained for $2000$ epochs.

The error distribution for the validation dataset is shown in \cref{fig:TCErrorHist}.
The results show that most errors are within about $2\%$. 
For $T_C$, $55,501$ of $56,711$ samples ($97.9\%$) have errors within $2\%$.
For $T_{\rm min}$, $56,095$ of $56,711$ samples ($98.9\%$) have errors within $2\%$.
The average absolute error of $T_C$ and $T_{\rm min}$ combined is $0.0095\;{\rm GeV}$, which is basically the same order of magnitude as the error range of the training data itself.

\subsection{Fitting of coefficients}

\subsubsection{Data preparation}

In general, the difference between two functions can be measured using,
\begin{equation} \label{eq:functiondistance} 
\varepsilon = \int_{\tau_0}^{\tau_1} d\tau \left( f_1(\tau) - f_2(\tau) \right)^2.
\end{equation}
This can be regarded as the Euclidean distance in a Hilbert space. 
To compare the differences between two functions using this metric for any $T_C$, the integration region should be unified.
Introducing $\tau = T/T_C$, the action can be transformed into $S_E(\tau) = \sum q'_i u^{(i)}\left(\tau\right)$, with $u^{(i)}(\tau)=\tau^i/(1-\tau)^2$ and $q'_i = T_C^{i-2} q_i $, where $q_i$ are the coefficients in Eq.~(\ref{eq:fit_fun}).

In practice, $T_{\rm min}$ is usually relatively large, so it is unnecessary to start the comparison from $\tau = 0$.
Meanwhile, the integral diverges at $\tau = 1$, making it impossible to extend the upper bound to $\tau \to 1$.
Therefore, the comparison between functions is carried out within a range $\tau_{\rm min} < \tau < \tau_{\rm max}$. 
From the statistical analysis of $T_{\rm min}$, it is found that $T_{\rm min}/T_C = 0.76 \pm 0.21$, so $\tau_{\rm min} = 0.55$ and $\tau_{\rm max} = 0.95$ are selected.
$\tau_{\rm min}$ and $\tau_{\rm max}$ are tunable hyperparameters that can be adjusted to optimize performance. 

An orthogonal function set is constructed by subtracting the projection of the original function onto the existing orthogonal basis, following the Gram-Schmidt orthogonalization process~\cite{10.1137/1.9780898718003}:
\[
w^{(i)}(\tau) = \frac{1}{N_i} \left[u^{(i)}(\tau)-\sum _{j=0}^{i-1}\left(\int_{\tau_{\rm min}}^{\tau_{\rm max}} d\tau' w^{(j)}(\tau') u^{(i)}(\tau') \right) w^{(j)}(\tau)\right],
\]
where $N_i$ is the normalization coefficient for $w^{(i)}(\tau)$. 
It can be verified that the orthogonal condition holds:
\[
\int_{\tau_{min}}^{\tau_{max}} d\tau w^{(i)}(\tau) w^{(j)}(\tau) = \delta_{ij},
\]
where $\delta_{ij}$ is the Kronecker delta.

In matrix form, this orthogonalization process can be expressed as a linear transformation.
Denoting $S_E(\tau)=\sum _i {q''}_i w^{(i)}(\tau)$, the coefficients ${q''}_i$ can be obtained as $\vec{q''}= (M^T)^{-1}\vec{q'}$, where $\vec{q''}=({q''}_0,{q''}_1,\ldots)^T$, $\vec{q'}=({q'}_0,{q'}_1,\ldots)^T$, $M=\left(d_0^T, d_1^T, \ldots \right)^T$ is a matrix can be obtained numerically, with $\vec{d}_i$ the vectors which can be calculated as:
\[
\vec{d}_i = \frac{1}{N_i} \vec{e}_i - \sum_{j < i} \frac{w_{ji}}{N_i} \vec{d}_j, 
\]
where $\vec{e}_i$ are unit row vectors (e.g., $\vec{e}_0 = (1, 0, 0, \dots)^T$, $\vec{e}_1 = (0, 1, 0, \dots)^T$, and $w_{ij} = \int_{\tau_{\rm min}}^{\tau_{\rm max}} d\tau w^{(i)}(\tau) u^{(j)}(\tau)$. 
With $n_{\rm order}=6$, $M$ is a $7\times 7$ matrix.
Since $u^{(i)}(x)$ are known, using $\tau_{\rm min}=0.55$ and $\tau_{\rm max}=0.95$, $w^{(i)}(\tau)$, $w_{ij}$ and $N_i$ can be calculated.
Then, the difference of two functions $S^a_E(\tau)$ and $S^b_E(\tau)$ in the region $[\tau_{\rm min}, \tau_{\rm max}]$ defined in Eq.~(\ref{eq:functiondistance}) can be expressed as $\varepsilon = |\vec{q''}^a - \vec{q''}^b|^2$.
Once $T_C$ is known and $\tau_{\rm min}$, $\tau_{\rm max}$ are chosen, one can calculate $q_i$ by using ${q''}_i$, and vice versa.
Therefore, in the following, the networks are trained to predict $\vec{q''}$.

\begin{figure}[thbp!]
\centering
\includegraphics[width=0.49\hsize]{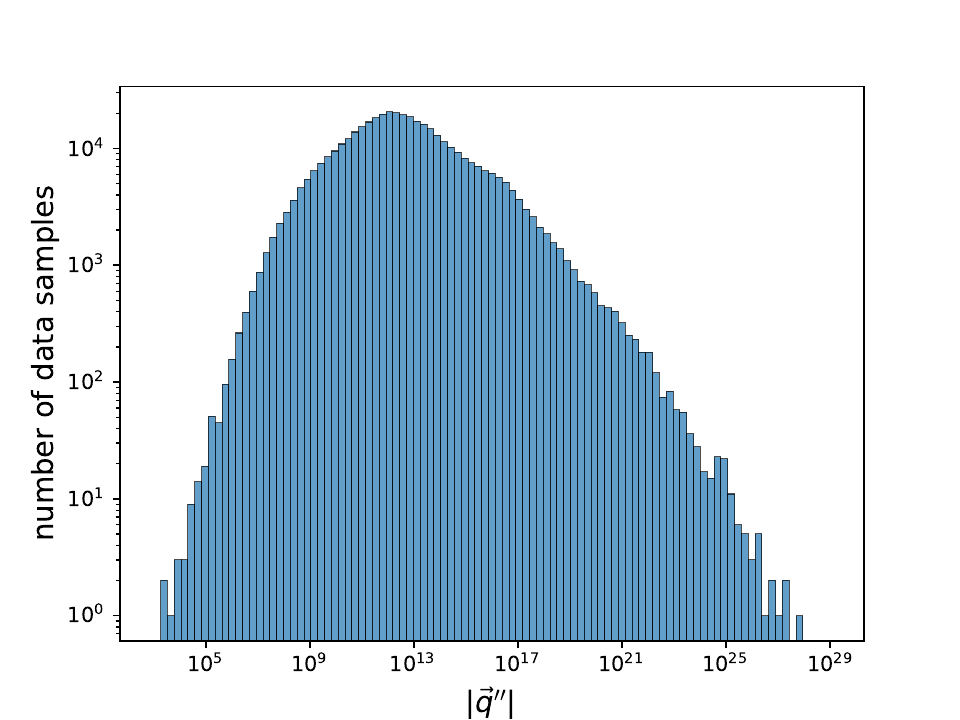}
\includegraphics[width=0.49\hsize]{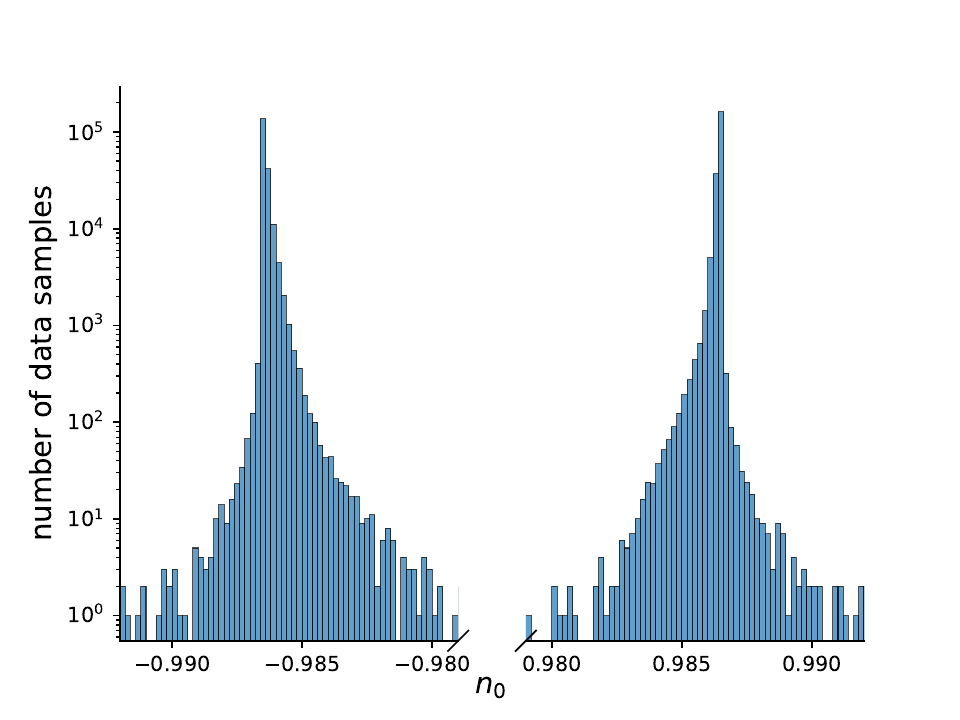}
\caption{The distribution of $|\vec{q''}|$~(the left panel) and $n_0$~(the right panel) over the whole dataset.
}\label{fig:qppdistribution}
\end{figure}
A major challenge in direct training is the large order-of-magnitude difference between the coefficients to be fitted. 
After constructing the orthogonal function set, this issue transforms into an order-of-magnitude difference in ${q''}_i$. 
The distribution of $|\vec{q''}|$ is shown in the left panel of \cref{fig:qppdistribution}.
The values of $|\vec{q''}|$ span a range of magnitudes from $10^3$ to $10^29$.
This is mainly because the $T^6$ term dominates the action $S_E(T)$: since most values of $T_C$ and $T_{\rm min}$ are greater than $1$, $T^6$ becomes very large, and the inherent large magnitude of $q_6$ further amplifies this effect. 
This leads to the directions of the $\vec{q''}$ vectors being very close, with the main difference lying in the magnitude of $|\vec{q''}|$.
Due to the extreme differences in the orders of magnitude of $|\vec{q''}|$, and the presence of a small number of outliers whose $|\vec{q''}|$ values differ significantly from the majority of the dataset, we filtered the dataset to prevent these points from affecting the training of the networks.

Another issue affecting data normalization is the distribution of ${q''}_i$. 
In the following, we define $\vec{n}=\vec{q''}/|\vec{q''}|$ as the direction of $\vec{q''}$. 
Taking $n_0$ as an example, the distribution of $n_0$ is shown in the right panel of \cref{fig:qppdistribution}. 
It can be observed that the vast majority of $n_0$ values are concentrated within two narrow peaks near $\pm 0.985$. 
This makes the z-score standardization ineffective for dispersing the data. 
However, $n_i$ exhibits an interesting property: for all $n_i$ in the dataset, almost all their signs are entirely consistent. 
Furthermore, if we consider only the absolute values of $n_i$, the distribution of $|n_i|$ is concentrated and suitable for z-score standardization. 
Therefore, we adopted an approach that first determines the sign of ${q''}_0$ and then fits $|n_i|$. 
Specifically, the sign of ${q''}_0$ is treated as a binary classification problem and predicted using one network, while $|n_i|$ are predicted as regression problems.

\begin{table}[htbp]
\centering
\begin{tabular}{cccc}
\hline
\hline
$|\vec{q''}|$ & $|n_0|$ & $|n_1|$ & $|n_2|$ \\
$<8\times 10^{18}$ & $[0.984,0.987]$ & $[0.15, 0.18]$ & $[0.035, 0.05]$ \\
\hline
$|n_3|$ & $|n_4|$ & $|n_5|$ & $|n_6|$ \\
$[0.005, 0.009]$ & $[0.0005, 0.001]$ & $[10^{-5}, 6\times 10^{-5}]$ & $[0.5\times 10^{-6}, 1.5\times 10^{-6}]$ \\
\hline
\hline
\end{tabular}
\caption{To prevent outliers from adversely affecting the training performance, the screening criteria are compiled.}
\label{tab:datafiltering}
\end{table}
The filtering conditions used to mitigate the impact of outliers on training are listed in \cref{tab:datafiltering}.
After filtering, $401,064$ samples remain out of the original $408,325$, with a retention rate of $98.2\%$.
The vast majority of data points been filtered out are removed due to excessively large $|\vec{q''}|$ values.
Before training, $|n_i|$ are standardized using the z-score standardization.
The standardization of $|\vec{q''}|$ is closely related to the loss function.
As will be introduced, in the training of the network to predict $|\vec{q''}|$, a special loss function is used.
To cooperate with the loss function, the $|\vec{q''}|$ is simply standardized by multiplying a factor $10^{-14}$.

\subsubsection{Structure of the networks}

In previous subsection, a network to predict $T_C$ and $T_{\rm min}$ is trained.
This network also serves as a part of the networks to predict the coefficients~(denoted as the `T' network.)
In the following, $T_C$ and $T_{\rm min}$ are used as known inputs to train the other networks.

As has been explained, analysis of the dataset shows that the signs of coefficients ${q''}_i$ are consistent—either all positive or all negative. 
As long as the sign is fitted correctly, with $|n_0| \approx \pm 0.985$, the direction of the $\vec{q''}$ can be roughly determined. 
Therefore, a classification task is first performed to distinguish between ${q''}_0>0$ and ${q''}_0 < 0$.
Such a network is denoted as the `sign decision' network.
The network is a densely connected network with $6$ input neurons~(the inputs are $c$, $m^2$, $\kappa$, $\lambda$, $T_C$, and $T_{\rm min}$), three hidden layers with $1024$ neurons, and an output layer with two neurons.
The two neurons serve as the activation units for the binary classification.
The PReLU function is used as the activation function between each layers.
On the output neurons, a `softmax' function is applied.
The loss function is chosen as the cross entropy.

After the `sign decision' network, seven networks are trained independently to predict the directions $|n_i|$, which are denoted as the `direction' networks.
Each of these networks are densely connected network with $7$ input neurons~(the inputs are $c$, $m^2$, $\kappa$, $\lambda$, $T_C$, $T_{\rm min}$, and ${\rm sign}({q''}_0)$), four hidden layers with $1024$ neurons, and an output layer with one neuron.
The PReLU function is used as the activation function.
The loss function is chosen as the mean squared error~(mse) function.

Finally, a KAN is used to predict $|\vec{q''}|$, which is denoted as the `length' network.
We adopt the \code{fast KAN} implementation~\cite{li2024kolmogorovarnold}.
The network is a densely connected network with $13$ input neurons~(the inputs are $c$, $m^2$, $\kappa$, $\lambda$, $T_C$, $T_{\rm min}$, ${\rm sign}({q''}_0)$, and $|n_i|$), five hidden layers with $64$ neurons, and an output layer with one neuron.
Due to the fact that $|\vec{q''}|$ is highly concentrated at relatively small values while there exist outliers with several orders of magnitude large values, we employed a special loss function to prevent these anomalous data points from adversely affecting the training performance:
\begin{equation} \label{eq:lossfunction} 
\text{Loss} = \frac{1}{\rho N} \sum \log (1+\xi^2(|\vec{q''}|^{\rm pred}-|\vec{q''}|)^2),
\end{equation}
where $N$ is the number of samples in the dataset, $\rho$ and $\xi$ are two tunable parameters, $|\vec{q''}|^{\rm pred}$ and $|\vec{q''}|$ are predictions of $|\vec{q''}|$ and the truths of $|\vec{q''}|$, respectively. 
Regions with larger slopes of the loss function can be considered as priorities when the model cannot fit all samples perfectly.
When $|\vec{q''}|^{\rm pred}-|\vec{q''}|\to 0$, $ \log (1+\xi^2(|\vec{q''}|^{\rm pred}-|\vec{q''}|)^2)/\rho \approx \xi^2 (|\vec{q''}|^{\rm pred}-|\vec{q''}|)^2/\rho$, so Eq.~(\ref{eq:lossfunction} ) can be regarded as an mse function multiplied by a constant $\xi^2/\rho$. 
For a very small $\left||\vec{q''}|^{\rm pred}-|\vec{q''}|\right|$, the training is considered satisfactory, so the slope is small (consistent with the mse loss function behavior).
For a very large $\left||\vec{q''}|^{\rm pred}-|\vec{q''}|\right|$, these samples are likely those with $|\vec{q''}|$ differing by several orders of magnitude, where small absolute errors are not required (their relative errors are already small), so they are de-emphasized.
In this work, $\rho = 10$ and $\xi = 100$ are used, as a consequence the priority is given to samples with errors around $0.01$.
The position of this priority region can be adjusted using $\rho$ and $\xi$, i.e., $\xi$ controls the position of the slope peak, and $\rho$ serves as a standardization of the loss function.

\begin{figure}[thbp!]
    \centering
    \includegraphics[width=0.99\hsize]{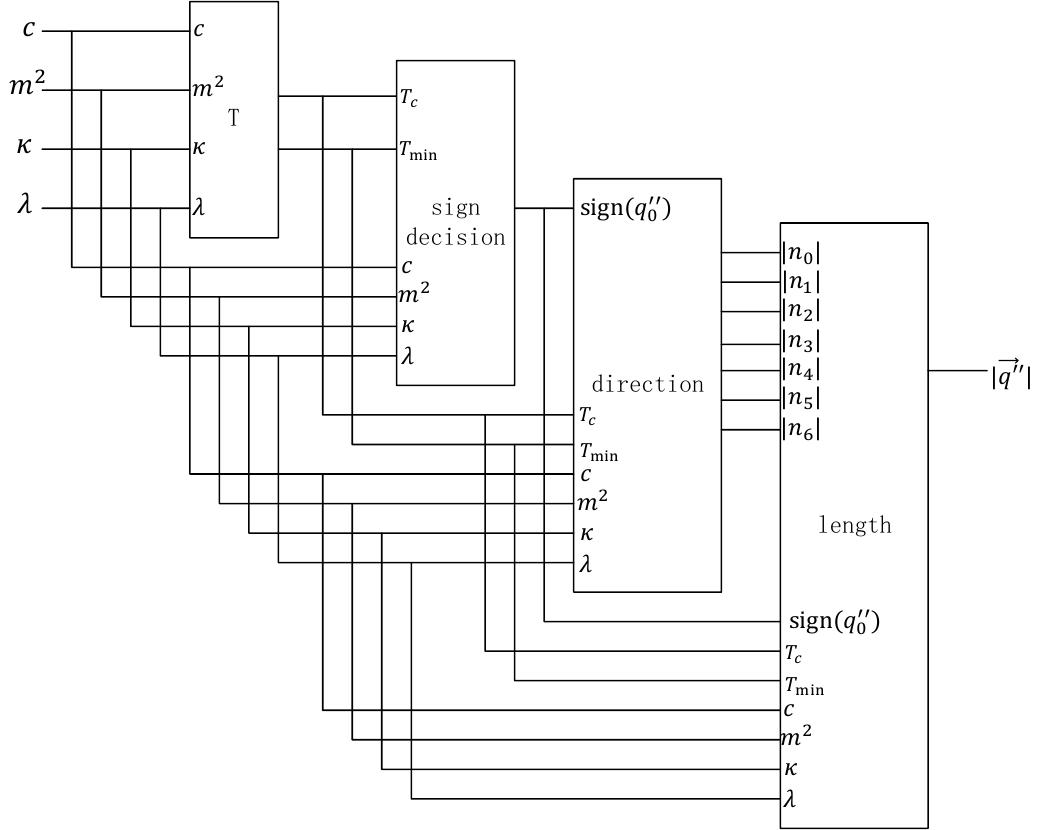}
    \caption{A sketch of the networks for predicting coefficients. 
A total of $10$ neural networks are used to predict the coefficients.
The first network predicts $T_C$ and $T_{\rm min}$ from $c$, $m^2$, $\kappa$ and $\lambda$. 
This is the network from the previous subsection, labeled as the `T' network in the diagram.
The second network predicts the sign of the coefficient ${q''}_0$ from $T_C$, $T_{\rm min}$, $c$, $m^2$, $\kappa$ and $\lambda$, and is referred to as the `sign decision' network.
Seven networks are used to predict the components of the direction $\vec{n}$~($|n_i|$) from ${\rm sign}({q''}_0)$, $T_C$, $T_{\rm min}$, $c$, $m^2$, $\kappa$ and $\lambda$, and are called the `direction' networks.
The last network predicts the length of the vector $\vec{q''}$ from $|n_i|$, ${\rm sign}({q''}_0)$, $T_C$, $T_{\rm min}$, $c$, $m^2$, $\kappa$ and $\lambda$, and is named the `length' network.
Then, the coefficients ${q''}_i$~(as well as $q_i$) can be obtained from $T_C$, ${\rm sign}({q''}_0)$, $|n_i|$ and $|\vec{q''}|$.
    }\label{fig:networks}
\end{figure}

A summary of the networks to predict the coefficients $q_i$ is shown in Fig.~\ref{fig:networks}.
During training, the inputs to the input layer are all sourced from the training set. 
Taking the `sign decision' network as an example, the $T_C$ and $T_{\rm min}$ required for training the `sign decision' network are obtained from the training set.
When predicting the coefficients $q_i$, the process starts from $c$, $m^2$, $\kappa$, and $\lambda$. 
Specifically, these parameters ($c$, $m^2$, $\kappa$, and $\lambda$) are first input to the 'T' network to obtain $T_C$ and $T_{\rm min}$. 
Then, $c$, $m^2$, $\kappa$, $\lambda$, along with the derived $T_C$ and $T_{\rm min}$, are fed into the `sign decision' network to predict ${\rm sign}({q''}_0)$. 
This process continues sequentially for each subsequent networks.

\subsubsection{Numerical results}

To train the `sign decision' network, the training set consists of $350,000$ data samples, while the validation set contains $51,064$. 
Around $100$ epochs, the cross entropy on the validation set stopped decreasing. 
After training, when using the $T_C$ and $T_{\rm min}$ predicted by the `T' network, $49,354$ samples ($96.65\%$) correctly predicted the sign of ${q''}_0$.

\begin{figure}[thbp!]
\centering
\includegraphics[width=0.6\hsize]{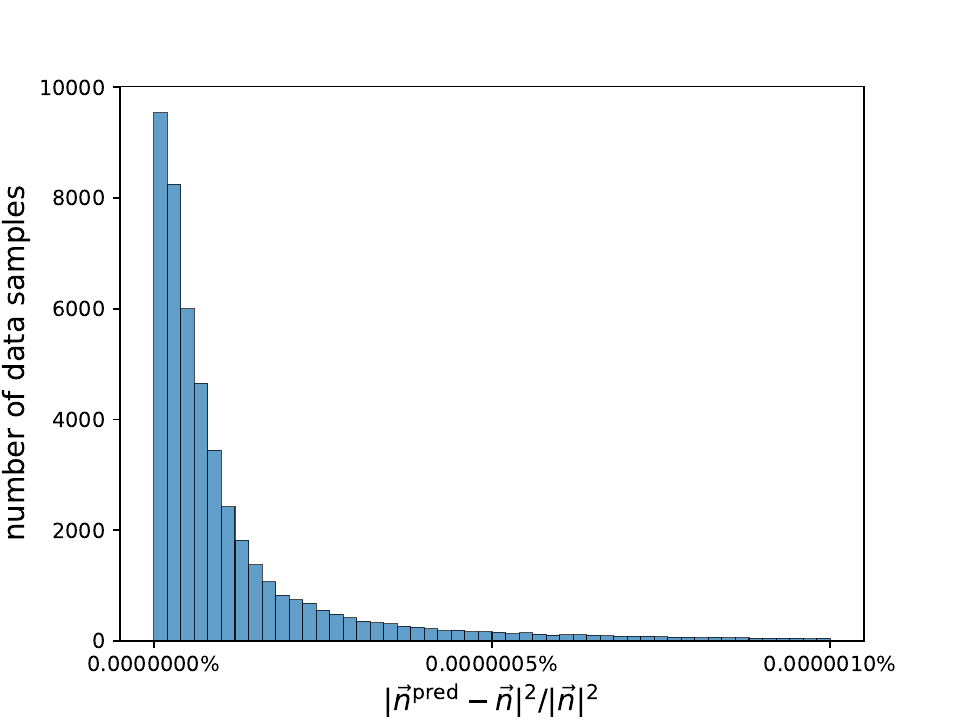}
\caption{Distribution of prediction errors for $\vec{n}$.
    }\label{fig:CoeffNormErrorHist}
\end{figure}

With the same training dataset and validation dataset, seven independent `direction' networks are trained (one for each $|n_i|$). 
Employing the learning curve from the validation set and early stopping to detect overfitting, the training was run for $1000$ epochs.
The error distribution is shown in \cref{fig:CoeffNormErrorHist}. 
It can be found that $96.59\%$ of the samples have a relative squared error within $10^{-5}$. 
This suggests that as long as ${\rm sign}({q''}_0)$ is predicted correctly ($96.65\%$), $|n_i|$ can be predicted accurately.

\begin{figure}[thbp!]
\centering
\includegraphics[width=0.49\hsize]{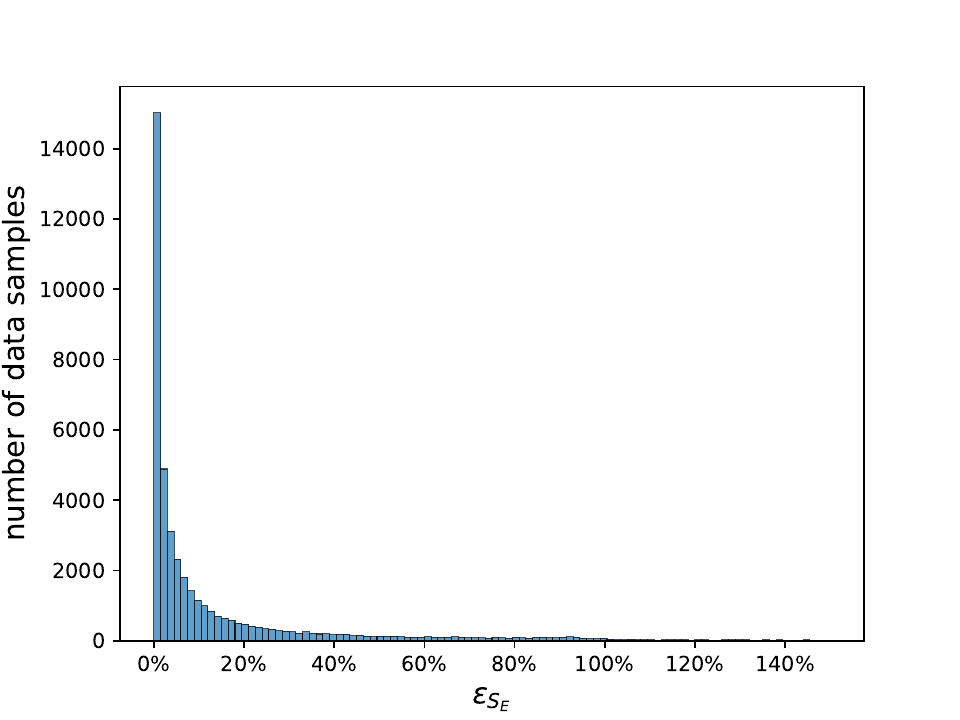}
\includegraphics[width=0.49\hsize]{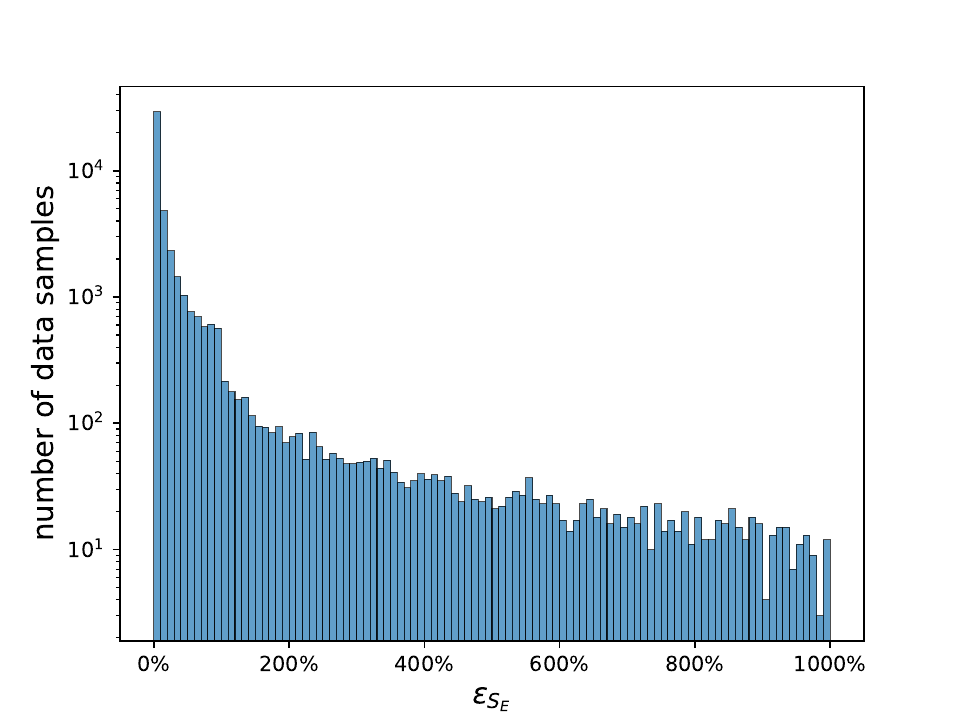}
\caption{The distribution of the relative errors $\varepsilon_{S_E}$ defined in Eq.~(\ref{eq:errordefinitionofse}).
The left panel shows the distribution of $\varepsilon_{S_E}$ confined within about $100\%$. 
The right panel displays the results over a broader range, where a long tail can be observed in the large-error region.
}\label{fig:reletiveErrorOfAction}
\end{figure}

During training, the most challenging part is the training of the `length' network. Throughout the process, the validation loss exhibited severe oscillations. 
Training is continued until $30,000$ epochs.
To quantify the error, we defined the relative error as,
\begin{equation} \label{eq:errordefinitionofse} 
\varepsilon_{S_E}=\frac{\int _{0.55T_C}^{0.95T_C}dT\left(S_E^{\rm pred}(T)-S_E(T)\right)^2}{\int _{0.55T_C}^{0.95T_C}dT S^2_E(T)},
\end{equation}
where $S_E^{\rm pred}$ is the action predicted by the networks. 
For the validation dataset, the relative error is shown in \cref{fig:reletiveErrorOfAction}. 
Note that, there is a long tail, which is not fully displayed in the left panel of \cref{fig:reletiveErrorOfAction}, i.e., there exists a small number of data points with large relative errors form a long tail close to the $\varepsilon _{S_E}$-axis as shown in the right panel of \cref{fig:reletiveErrorOfAction}. 
This long tail is a result of the special loss function we employed.
Results on the validation set show that for $S_E(T)$ within the range of $0.55T_C$ to $0.95T_C$, approximately $71.48\%$ of the data points have a relative error of less than $30\%$.

It can be observed that identifiable factors affecting the fitting results include: a tendency for large errors to occur near the edge of the phase transition region where $|\kappa|$ is relatively small, as well as when $m^2\to 0$. 
Additionally, the dataset contains some instances where $T_{\rm min} > 0.95T_C$. 
Notably, a number of outliers with significant errors are concentrated in cases where $T_{min}$ is close to $T_C$ (in fact, while most data points with $T_{\rm min}$ near $T_C$ yield good fitting results, those with large fitting errors are also concentrated in this same region). 
Therefore, we believe that better results could potentially be achieved by further trimming the dataset and, as a compromise, focusing training and prediction only on a majority of the region of parameter space in interest, while deeming predictions unreliable outside this region.
In other words, to archive a better result, one can shrink the region of phase transition in the parameter space to define a `fittable' region, and deduce a training on this smaller `fittable' region.
This represents an avenue for future exploration.

\section{Conclusions}
\label{conclusion}



The calculation of the bounce action for phase transitions, which entails solving non-trivial partial differential equations, is inherently prone to numerical uncertainties. These uncertainties are further amplified when deriving key transition characteristics—such as the nucleation temperature, percolation temperature, and inverse transition duration—since their evaluation involves both differentiating and integrating the noisy action function.

In this work, we address this challenge by proposing a fitting approach that models the action as a smooth function of temperature. This strategy effectively mitigates the propagation of numerical errors in the determination of these quantities. Our analysis shows that, after factoring out a scaling term of $1/(T-T_C)^2$, a sixth-order polynomial provides an excellent fit to the action data within the high-temperature approximated potential. For a more realistic case, the scalar singlet model, this method also achieves satisfactory performance across most of the parameter space, provided that the input action data are appropriately preprocessed. The advantages of this approach can be summarized as follows:
\begin{itemize}
\item \textbf{Enhanced Accuracy:} By employing a smooth analytical representation of the action, the method substantially reduces numerical errors introduced during differentiation and integration, thereby yielding more reliable and precise determinations of phase transition properties.
\item \textbf{Improved Computational Efficiency:} The fitting procedure eliminates the need to compute the action at every temperature point, significantly decreasing the total computation time—an advantage that becomes particularly pronounced in extensive parameter-space scans.
\item \textbf{Facilitated Error Analysis:} The analytical nature of the fit provides a transparent framework for systematic error estimation and propagation, enhancing the statistical robustness of the results.
\item \textbf{Machine Learning Extension:} We further demonstrate that trained neural networks can directly predict the action curve from model parameters. This promising extension points toward a powerful tool for the rapid evaluation of phase transition dynamics.
\end{itemize}

We have added this method into the latest version of \code{PhaseTracer}, and also provided a code snap to perform action fitting in \code{CosmoTransitions}.

\addcontentsline{toc}{section}{Acknowledgments}
\acknowledgments
This work was supported by the National Natural Science Foundation of China (12335005, 12105248) and 
the PI Research Fund from Henan Normal University under Grant No. 5101029470335. L.B. is supported by the National Natural Science Foundation of China (NSFC) under Grants Nos. 12322505, 12347101, L.B. also acknowledge Chongqing Talents: Exceptional Young Talents Project No. cstc2024ycjh-bgzxm0020 and Chongqing Natural Science Foundation under Grant No. CSTB2024NSCQ-JQX0022. J.-C. Y. is supported in part by the National Natural Science Foundation of China under Grants Nos. 12147214 and 12575106, and by the Natural Science Foundation of the Liaoning Scientific
Committee No. LJ212510165024.

\appendix

\section{Fit action curve in \code{PhaseTracer}}

We have added the action fitting in the latest version of \code{PhaseTracer2}. It is implemented within the TransitionFinder class and can be enabled using the 
\begin{lstlisting}
  ... ...
  // Make TransitionFinder object and find the transitions
  PhaseTracer::TransitionFinder tf(pf, ac);
  tf.set_fit_action_curve(true);
  tf.find_transitions();
\end{lstlisting}
When \code{fit\_action\_curve} is set to true, the algorithm first checks if the available temperature range for fitting exceeds 2 GeV. If this condition is met, it generates \code{action\_curve\_nodes} points uniformly across this temperature range. It then applies the selection criteria outlined in Eq. (22) to these points. A fit is only attempted if the number of valid points remaining after this selection is greater than $2*$\code{action\_curve\_order}.

Should any of the above checks fail, or if the resulting fit has a RMSE greater than 5, the algorithm will revert to the traditional method for calculating the transition parameters. If the fit is successful, the resulting action curve is used to calculate the $T_N$ and  $\beta$. A sample output demonstrating a successful fit is shown below:

\begin{lstlisting}
=== transition from phase 0 to phase 1 ===
changed = [true] 
TC = 59.1608
false vacuum (TC) = [-6.18626e-06]
true vacuum (TC) = [50.0003]
gamma (TC) = 0.845159
delta potential (TC) = 0.00117818
Action curve fitting succeeded with MSE = 0.00144928
TN = 57.4319
false vacuum (TN) = [-6.93692e-06]
true vacuum (TN) = [53.5322]
\end{lstlisting}

\section{Fit action curve in \code{CosmoTransitions}}

To use the action fit to calculate $\beta$ within \code{CosmoTransitions}, one can replace the \code{tunnelFromPhase} function in 
\code{transitionFinder.py} by the codes in \url{ https://github.com/phyzhangyang/Use_ActionFit_in_CosmoTransitions}
After that, it can be called using 

To calculate $\beta$ using the action fitting method within \code{CosmoTransitions}, one can replace the \code{tunnelFromPhase} function in \code{transitionFinder.py} with the implementation provided in the repository \url{ https://github.com/phyzhangyang/Use_ActionFit_in_CosmoTransitions}. 
After implementing this modification, the functionality can be invoked using the following code:
\begin{lstlisting}
  m = model1()
  m.findAllTransitions()
    for tran in m.TnTrans:
      print('Tnuc=',tran['Tnuc'])
      print('action=',tran['action'])
      if 'd(S/T)/dT' in tran:
        print('d(S/T)/dT=',tran['d(S/T)/dT'])
\end{lstlisting}

\addcontentsline{toc}{section}{References}
\bibliographystyle{JHEP}
\bibliography{ref}
\end{document}